\documentstyle[aasms4]{article}
\def \ni {\noindent}

\def \gta {\mathrel{\vcenter 
     {\hbox{$>$}\nointerlineskip\hbox{$\sim$}}}} 
\begin{document}

\title{The Galactic Shock Pump: A Source of Supersonic Internal Motions in the Cool
Interstellar Medium}

\author{Paul Kornreich and John Scalo}
\affil{Astronomy Department, University of Texas, Austin, TX 78712}
\begin{center}
{{\it The Astrophysical Journal}, in Press}\\
\end{center}
 
\begin{abstract}
We propose that galactic shocks propagating through interstellar density 
fluctuations provide a mechanism for the intermittent replenishment, or ``pumping," of the
supersonic motions and internal density enhancements observed pervasively within cool atomic
and molecular interstellar structures, without necessarily requiring the presence of
self-gravity, magnetic fields, or young stars.  The shocks are assumed to be due to a variety
of galactic sources on a range of scales.  An analytic result for the kinematic vorticity
generated by a shock passing through a radially-stratified two-dimensional isobaric model cloud is derived,
assuming that the Mach number is not so large that the cloud is disrupted, and neglecting the shock curvature
and cloud distortion.  Two-dimensional lattice gas hydrodynamic simulations at modest Mach numbers were used
to verify the analytic result.
The induced internal velocities are initially a significant fraction of the shock speed divided by the
square root of the density contrast, accounting for both the observed linewidth amplitudes and the
apparent cloud-to-cloud linewidth-density scaling.  The linewidth-size relation could then be
interpreted in terms of the well-known power spectrum of a system of shocks.  The induced
vortical energy should quickly be converted to compressible and MHD modes, and so would be difficult to
observe directly, even though it would still be the power source for the other modes.  The shockpump
thus produces density structure without the necessity of any sort of instability.  We argue that the
shockpump should lead to nested shock-induced structures, providing a cascade mechanism for supersonic
``turbulence" and a physical explanation for the fractal-like structure of the cool interstellar
medium.  The average time between shock exposures for an idealized cloud in our Galaxy is estimated and
found to be small enough that the shockpump is capable of sustaining the supersonic motions against
readjustment and dissipation, except for the smallest structures.  This suggests an explanation of the
roughly spatially uniform and nearly sonic linewidths in small ``dense cores."  We speculate that the
avoidance of shock pumping may be necessary for a localized region to form stars, and that the inverse
dependence of probability of avoidance on region size may be an important factor in determining the
stellar initial mass function.
\end{abstract} 
 
\section{INTRODUCTION}
\subsection{Motivation}
	The source of the supersonic linewidths observed within interstellar density fluctuations, 
or ``clouds," remains obscure.  Supersonic internal velocities are observed in a wide
variety of structures covering an enormous range of densities and sizes, from over 1000 pc
down to 0.1 pc and smaller.  Often the velocity dispersion $\sigma$ of the unresolved
structure, as measured by the spectral linewidth, is found to scale with region size $\ell$
as a power law, 
$\sigma\sim\ell^\alpha$ , with $\alpha\approx0.4-0.6$ (e.g. Larson 1981, Falgarone, Puget,
\& Perault 1992, Myers \& Goodman 1988a,b, Fuller \& Myers 1992, Elmegreen \& Falgarone
1996), although the situation is not so clear.  For example, the correlation seems weaker,
shallower, or non-existent in regions of vigorous high-mass star formation (Caselli \& Myers
1995, Plume et al. 1997), and no correlation of linewidth with size was found in a
homogeneous C$^{18}$O study of 40 low-mass cores by Onishi et al. (1996), although the
dynamic range in size was small.  The latter study did find a correlation of linewidth with
column density.  Xie (1997) has suggested that it is the linewidth-density anti-correlation
that may be more physically fundamental, a point to which we return below.  It is also
important to realize that these scaling relations may refer to different types of samples and
tracers (Barranco \& Goodman 1998).  Here we are referring to region-to-region, or
``multicloud" correlations, not scaling within individual poorly resolved structures.

These supersonic motions have been variously 
attributed to some form of compressible ``turbulence," virialized motions induced by
self-gravity, inverse angular momentum cascades, MHD waves, and protostellar winds (for recent
theoretical discussions and references see McLaughlin \& Pudritz 1997, Vazquez-Semadeni,
Ballesteros-Paredes \& Rodriguez 1997, Xie 1997; see Scalo 1987 for a review of other
suggestions), but none of these seem compelling for a number of reasons.  First, supersonic
linewidths have been known since the 1960s to exist in diffuse HI clouds, which are not
self-gravitating and have no embedded protostars.  The empirical scaling relations for
these regions may be similar to the relations for self-gravitating clouds (Quiroga 1983; 
see also Fleck 1996).  Recent work (Heithausen 1996) indicates that diffuse non-star-forming
molecular clouds have the same scaling laws as for self-gravitating clouds, although the
inferred scaling is uncertain because of distance uncertainties.  In any case, it is firmly
established that most clouds, atomic or molecular, in which self-gravity is unimportant have
strongly supersonic linewidths; see, for example, the study of MBM12 by Pound, Bania, \&
Wilson (1992), the study of MBM7 by Minh et al. (1996), and the survey by Heithausen (1996). 
Most of the derivations of the scaling relations assume virialization (e.g. de Vega,
Sanchez, \& Combes 1995).  Besides not explaining how virialization can occur or be
sustained without a power source such as protostellar winds, such models have the obvious
deficiency of not accounting for supersonic linewidths in non-self-gravitating regions. External
pressure confinement could be arbitrarily invoked for these cases, but the physical basis for the
concept of cloud pressure confinement, whether thermal or turbulent, seems doubtful for a
supersonically turbulent ISM, as recently demonstrated using numerical simulations by
Ballesteros-Paredes, Vazquez-Semadeni, \& Scalo (1999).  

Some proposals involve other requirements that are difficult to
justify theoretically or observationally.  For example, in order to match the ``standard"
(but uncertain) scaling relations, MHD wave models require (besides the assumption of
virialization which would not apply to the diffuse clouds) that the fluctuation amplitude be
independent of size scale, a requirement that not only has no observational basis, but is
contradicted by two independent simulation studies (see the discussion in Xie 1997).  Other
problems with MHD waves as the source of the linewidths, based on simulations, have
been given by Stone (1994), Padoan \& Nordlund (1999), Mac Low et al. (1998), and Ostriker 
et al. (1998). 

Finally, and most relevant for the
present work, all the proposed mechanisms suffer a ``decay problem," in that the motions
should dissipate in a time which is comparable to the cloud crossing time, unless a power
source is invoked. MHD wave damping may allow a somewhat longer, but uncertain, dissipation
time, but still requires an initial power source, as well as possible replenishment. 
Pudritz (1995) and Stone (1994) have pointed out that the source of excitation of 
nonlinear MHD waves is unknown.  In particular, Stone (1994) showed by numerical simulations
that the highly tangled fields required to support a cloud are rapidly dissipated by
reconnection, so a replenishing source of energy is required.  (For another view of the
possible effects of reconnection, see Lubow \& Pringle 1996.)  Recent simulations by Mac
Low et al. (1998), Ostriker et al. (1998), and Mac Low (1999) result in similar temporal
energy decay for MHD and non-MHD compressible turbulence, leading these authors to also
conclude that an external energy source is required.  

All of this suggests that
the physical processes associated with the observed linewidths  require a continuously-available energy
source which does not depend on the importance of self-gravity or internal protostars, although
certainly these effects may modify the resulting behavior or even dominate in some cases. One such
source would be galactic rotation (Fleck 1981).  However it has proven difficult to understand how to
couple galactic rotation to internal cloud motion.  Das \& Jog (1995) investigated the
heating due to the periodic variation of the galactic tidal field across a cloud, but the
resulting heating rate was far too small, at least for clouds in a galactic disk.  

In the present paper we suggest that interstellar shocks, driven by galactic star 
formation on all scales from superbubbles down to protostellar winds (see Norman \& Ferrara
1997 for a calculation of the broad-band source spectrum), or by the supersonic turbulence itself, can
provide such an energy source.  Specifically, we show that the internal motions initially induced by the
passage of a shock through a cloud with an internal density gradient will generate vortical motions
which scale inversely as the square root of the cloud density contrast, that the corresponding compressible and
bulk cloud motions from such shocks will be comparable in magnitude and possess the same
density scaling, and that the frequency of such shock-cloud encounters is sufficient to keep
most clouds ``pumped" with internal motions.
	
\subsection{Previous work}
	Given the acknowledged role played by stellar explosions and winds in the large-scale, global, energy
balance of the ISM, it is perhaps surprising that shock-induced motions, originating outside
a given region, have not been previously investigated as a source for the supersonic
linewidths  {\it within} cool dense regions of the ISM.  McCray and Snow (1979) summarized work
that emphasized the role of galactic shocks in accounting for the hot and warm thermal
``phases" of the ISM (see also Clifford 1984).  Several authors (e.g. Zel'dovich, Ruzmaikin,
and Sokoloff 1983, sec.VI.3) have shown that stellar-driven energy input may be sufficient to
account for the overall HI velocity dispersion and scale height in the face of dissipation,
and Miesch \& Bally (1994) presented a similar energy balance argument for the interiors of
GMCs (see also Norman \& Silk 1980 for a theoretical model).  Norman \& Ferrara (1997)
examined the consequences of an assumed equilibrium between broad-band stellar energy input
and radiative cooling for a nonlinear transfer function appropriate to incompressible
turbulence, and calculated the scale-dependent source spectrum of several stellar energy injection
processes.  However here we are not examining the global energy budget on any scale, but are instead
concerned with a specific mechanism by which shocks can transfer their ordered kinetic energy into
``turbulent" motions within a localized density fluctuation, or ``cloud," in order to account for the
observed supersonic linewidths.  The only directly related study we know of is the numerical
simulations of Keto \& Lattanzio (1989), who suggested that cloud-cloud collisions could account for
the supersonic linewidths in the diffuse high-latitude molecular clouds.  More recent
highly-resolved MHD simulations by Miniati et al.(1999) in two dimensions and
especially Klein \& Woods (1998) in three dimensions show vividly how cloud-cloud
collisions should generate small-scale substructure.  In the present work we are mostly
concerned with shocks that originate from external sources, although cloud collisions
should have a similar effect.

 Miesch \& Zweibel (1994) identified four ways in
which shock energy may be imparted to the  interstellar medium.  1. The dense shells that
form behind expanding shock waves due to radiative cooling create material moving at the
shell velocity; 2. Asymmetric heating of a cloud near hot stars can accelerate the cloud due
to the ``rocket effect"; 3. A passing shock can directly accelerate the cloud, resulting in a
``bulk velocity"; 4. Shocks may generate a spectrum of waves.  Mechanisms 1-3 accelerate the
cloud as a whole, and do not involve internal motions, although they may certainly contribute
to the observed velocity distribution of interstellar structures.  The wave generation is
considered to be due to reflection from the cloud boundary when the ``cloud" is imagined as
an object with a sharp edge.  On the contrary, the present work is concerned with how
internal motions and structure can be generated by shocks passing through clouds with
continuous internal density gradients.
	
Most previous studies of shocks propagating into inhomogeneous astrophysical media have 
been numerical, and have focussed on interactions between a shock and a sharp-boundaried 
cloud, a problem first studied with fairly high resolution by Woodward(1976).  The review by
McKee (1988) summarizes the main stages: creation of the reflected and transmitted shock;
cloud compression; re-expansion of the cloud; and the onset of Kelvin-Helmholtz \&
Rayleigh-Taylor instabilities, with the final fate determined by initial conditions. 
Detailed numerical simulations concentrating on cloud disruption at large Mach numbers have
been presented by Bedogni \& Woodward (1990), Klein et al. (1994), and MacLow et al. (1994)
in two dimensions and Stone \& Norman (1992) and Xu \& Stone (1995) in three dimensions.  Vanhala \&
Cameron (1998) presented three dimensional simulations, including a very detailed treatment of the
cooling rates, that addressed the question of disruption versus gravitational instability, for
relatively high shock speeds and relatively sharp-edged (Gaussian) clouds.  Horvath
\& Toth (1995) presented two-dimensinal simulations of lower-velocity non-disruptive shock-cloud
interactions. In all these simulations, at least for early times, the vorticity is present only as a
sheet on the outside of the cloud.  In contrast, the present work is analytical, applicable to the case
of a general {\it continuous} pre-shock density gradient, and concentrates on the {\it internal}
velocity field induced by non-disruptive shock-cloud interactions.  
	With regard to continuous density distributions, studies of shocks passing through a 
radially decreasing density distribution have been given by Picone et al. (1983) and Picone
\& Boris (1983, 1988), who demonstrated the development of two oppositely signed vortices
inside the two halves of the ``hole."  The physical situation we analyze below is similar to
this, but with an increasing density distribution.  Our analytic approach is also
completely different from that proposed in Picone et al.  Recently Foster and Boss (1996) have
simulated winds incident on self-gravitating clouds with internal density gradients, but
their work was concerned with the demonstration of the crucial role of the assumed adiabatic
index on whether  shocks would disrupt or instigate collapse in marginally stable clouds, and
so the internal motions were not discussed.  Schiano, Christiansen, \& Knerr (1995) simulated
high-speed winds incident on ram pressure-confined clouds, but it is not clear how stratified the
initial density distribution was, and they did not examine the induced internal velocity
fluctuations (see sec. 3.7 for a discussion of implications for the present work).  Kimura
\& Tosa (1993) and Elmegreen et al. (1995) studied the interaction of a shock with a
continuous density-velocity field set up by an initially random velocity field, but the
emphasis was on the shock corrugation and the irregular accumulation of mass behind the
shock.  
	None of these studies have considered the question of internal motions induced by the 
passage of a shock through inhomogeneous interstellar structures.  

\subsection{Outline of the present model}
In the present paper we show
that the vorticity ${\bf\nabla\times u}$ (and dilatation ${\bf\nabla\cdot u}$) induced in a
spherically symmetric model cloud with a radial density gradient will be concentrated within
the cloud, and provides a significant source of internal kinetic energy which is available to
power other modes of cloud motions (e.g. compressible modes or MHD waves).  We claim that the
possible fates of a cloud exposed to a shock are not limited to disruption or collapse, but
instead shocks may deform and ³stir up² a cloud, perhaps many times, before the cloud is
eventually disrupted, coalesces with other density fluctuations, or suffers some other sort of
evolutionary path, so the effects of the repeated shocking is in effect to keep the cloud
``pumped up" with internal kinetic energy.  For this reason we refer to the process as the
``shock pump."  The cloud may eventually disperse or collapse or lose its identity in some
other way, but for much of its life the source of its internal dynamical motions would be
interactions with shocks.  
	
Our general model picture of the interstellar medium is basically a field of 
shock waves generated by sources on a wide range of scales, and occurring within and between a
complicated density field that likewise has structure on all scales.  The shocks could represent
winds from low-mass protostars, ionization-shock fronts, supernova remnants, superbubbles powered
by clusters of stars, larger scale shocks generated by infalling gas, cloud collisions, or
supersonic turbulence at scales larger than the model cloud.  This picture of supersonic
turbulence as interacting fields of shocks and density inhomogeneities, both with structure
at all scales, and coupled by star formation, is a generalization of the picture that served
as the motivation for the unfinished project of von Hoerner, von Weizacker, and coworkers to
model interstellar turbulence as a field of shocks (see von Hoerner 1962).  However for the
purposes of the present calculation we are only considering a single interaction of an
idealized shock with a very idealized ``cloud."  The statistical description of a large
ensemble of such interactions on a range of scales is left to future work, although we
speculate below on how the mechanism described here could be at least part of the physics
behind the fractal-like structure of the cool ISM.
	
After this work was complete, we found that Chernin and collaborators (see Chernin 1996 
and references therein) have independently studied the problem of a shock incident on a cloud
with a continuous density gradient (as well as shock-shock interactions).  Although they were
mostly concerned with angular momentum generation in cosmogony, and even though their approach to
the problem is very different than presented here, there is fair agreement for the predicted
induced velocities and their spatial distribution, although the scaling they suggest, based
on a weak-strong scattering theory, does not agree with the results found here, mostly
because their results are only valid for the limit of very small density contrasts, as
discussed in sec. 3 below.

In sec. 2 we derive a fairly general expression (eq. 36) for the kinematic vorticity 
generated by a linear shock propagating into an inhomogeneous two-dimensional medium.
In sec. 3 we apply this result to the passage of a shock over a radially stratified cloud,
and derive the root-mean-square vorticity and velocity amplitude due to
the shock passage.  The induced bulk (i.e. center of mass/cloud) motion is also calculated,
and is shown to be approximately equal to the induced vortical and (by symmetry) compressional
motions.  We show that the induced velocity amplitudes should scale inversely with the
square root of the density.  The rate at which clouds should be ``pumped" by shocks is
estimated in sec. 4, where we conclude that, for a randomly chosen ``cloud," the average time
between shock encounters is smaller than the time for readjustment of the cloud, except for
the very smallest structures.  Some implications of the results for models of ``coherent
cores," the stellar initial mass spectrum, and the spatial organization of star formation in
galaxies are discussed in sec. 5.

\section{KINEMATIC SHOCK VORTICITY IN AN INHOMOGENEOUS MEDIUM}
\subsection{Description of the vorticity generation model}

When a planar shock propagates into a medium with an inhomogeneous density, vorticity is
created by the variation of the post-shock speed along the surface of the shock. For
density gradients at constant pressure, the sound speed ahead of the shock varies along the
shock.  If the Mach number does not vary much along the shock surface, the shock will slow
down in the higher-density regions.  For isothermal density inhomogeneities a similar effect
occurs except that it is due to a changing Mach number, instead of a changing sound speed. 
The result in both cases is that the normal component of the postshock speed will vary along
the shock surface; this gradient of the normal component of the postshock velocity in the
tangential (parallel to the shock surface) direction is a vortical mode.  In order to create
this vorticity directly by the shock, the density gradients must exist along the shock
surface.  The situation is illustrated in Figure 1, which shows the field of postshock
velocity vectors generated by a linear shock as it crosses a circularly symmetric centrally
condensed model cloud.  We show below that the induced vorticity depends on the characteristic
scale of the density gradient in the tangential direction,
$\delta^{-1}_\tau=(\partial \ell n\rho/\partial\tau)^{-1}$, where $\tau$ is coordinate 
measured along the shock surface.  

\begin{figure}
\plotone{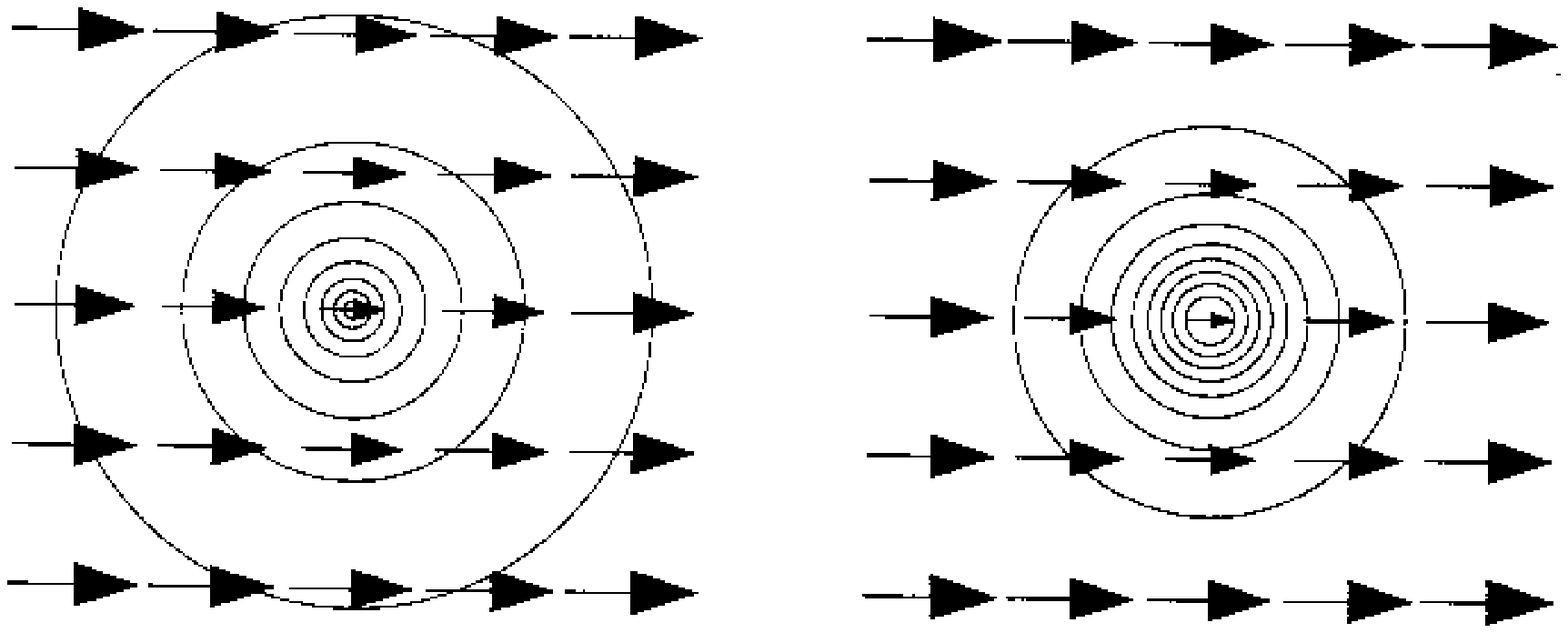}
\figcaption{Distribution of postshock velocity vectors for a linear shock passing
(from left to right) through a circular cloud, ignoring effects such as cloud compression or
shock curvature.  The size of each arrowhead is proportional to the postshock speed at that
position.  The density distributions are power laws with indices -1 (a) and -2 (b), and
density contours are indicated by the circular lines.  In this case the kinematic vorticity
is due to the derivative of the postshock velocity along the shock line.}
\end{figure}

Our model consists of a planar shock of speed $v_{sh}$, propagating in the x-direction,
incident on a spherically symmetric centrally condensed cloud whose density distribuition
$\rho(r)$ joins smoothly with the ambient density $\rho_a$ at large r.  For the purpose of
the present paper we assume that the shock remains planar as it propagates through the
cloud.  The inclusion of the shock curvature (see Fleck 1991 for an ISM application) would increase the
vorticity generation, and could be included iteratively using the formalism given here, but we
postpone a treatment of curvature effects for simplicity.  We also neglect the vorticity generated by
Mach stem and related phenomena (see Klein et al. 1994; Chernin 1996 and references therein).  For the
case of a planar shock, the vorticity generation is then due to two distinct processes.  The first
process is purely kinematical, and arises because the postshock velocity depends on the
preshock density, which varies radially through the cloud.  A gradient in the normal
component of the postshock velocity will then exist along the shock surface, which is the
source of vorticity.  We emphasize that this effect is purely kinematic in origin, and is not
related to the baroclinic vector $\nabla p\times\nabla\rho$ or the dilatation term
$\omega\nabla\cdot u$ in the vorticity equation, nor does it depend on the energy equation. 
As we show below, the process basically traces to the asymmetry between conservation of
normal and tangential momentum components at a discontinuity. For the simple model
considered here, the passage of the shock will also generate a gradient of the postshock
velocity perpendicular to the shock surface (i.e. along the x-axis) which is identical to the
vortical gradient, but rotated by
$\pi/2$.  In the uniform cloud density case this component would be identified with a
compression and acceleration of the cloud.

The second process results from the amplification or decay of the
kinematically-produced vorticity by the flow field further downstream in the postshock flow, due
to the baroclinic and dilatation terms in the vorticity equation.  Although we have included
this second process of ``dynamical vorticity production" in our full calculations (Kornreich \& Scalo
2000, in preparation, hereafter KS00), it is mathematically much more complicated, since it involves
the evaluation of the flow derivatives from the full hydrodynamic equations.  Our results (KS00) show
that the vorticity change due to this process is generally much smaller than the kinematic effect,
except for Mach numbers very close to unity, so we omit it here in order to give a more concise
presentation.

We are conceptualizing the model ``cloud" as imbedded in some uniform ``intercloud medium."  The same shocks
that we claim can drive dynamical motions within the cloud will also affect the intercloud gas, perhaps
sweeping it away from the cloud, or, if the intercloud medium is inhomogeneous (in which case it can be
regarded as other ``clouds"), generating vortical and compressible velocity fluctuations there.  Whatever the
nature of these intercloud flows, they will have large speeds because of the lower density.  However the
present paper is only concerned with the velocity fluctuations induced within the cloud, and we do not pursue
the interesting questions surrounding the fate of the intercloud gas.

In addition, it is likely that for some combinations of parameters, the shock-cloud interaction will lead
either to disruption of the cloud or gravitational instability by compression.  It is unclear what the shock
and cloud parameters should be to arrive at these end states, so it must be remembered that we are only
treating the cases in which the shock-cloud interaction leads to internal motions that do not disrupt the
cloud or cause it to collapse.  A very thorough simulation study of the collapse/dispersal question has been
presented by Vanhala \& Cameron (1998).  However they concentrated on relatively large shock velocities and
assumed cloud density distributions that were much more sharp-edge (e.g. Gaussian) than considered here.  The
fraction of parameter space occupied by our results is uncertain although we argue below that cloud
disruption should occur in a small fraction of cases.

\subsection{Outline of the derivation}
Our treatment of kinematic vorticity induced by shock-density fluctuation interactions 
parallels the studies of curved shock vorticity by Truesdall (1952, 1954) and Hayes
(1957).  For a discussion of the MHD curved shock problem, see Ram (1967) and Ram \&
Upadhyaya (1968) and references therein.  It is convenient to express the relevant equations
in a coordinate system whose unit vectors, $\hat n$ and $\hat\tau$, are normal and tangential
to the local shock surface.  This transformation is especially useful in generalizing our
results to include shock curvature and the presence of preshock velocity fluctuations
(KS00).  

It is useful to review the necessary transformations and derivatives relating to the
coordinate vectors.  In what follows, subscripts ijklm refer to cartesian coordinates, and,
unless otherwise specified, summation over repeated indices is implied.  A comma before a
subscript denotes partial differentiation, e.g. $v_{i,j}\equiv\partial v_i/\partial x_j$.  The
components of a vector $v_i$ in this coordinate system are simply the normal component
$v_n\equiv v_in_i$ and the tangential component $v_r\equiv v_i\tau_i$.  The normal component
may be separated out by the relation
\begin{equation}
v_i-v_nn_i=\epsilon_{ijk}\epsilon_{klm}n_jv_ln_m\ .
\end{equation}
\ni where $\epsilon_{ijk}$ is the usual alternating permutation symbol.
Since the normal and tangent vectors are gradients of the coordinates,
\begin{equation}
\epsilon_{ijk}n_{k,j}=\epsilon_{ijk}\tau_{k,j}=0.
\end{equation}
\ni Also, since they have fixed lengths,
\begin{equation}
n_{i,i}=\tau_{i,i}=0.
\end{equation}
\ni The directional derivatives are
\begin{equation}
n_i\nabla_i=\frac{\partial}{\partial n},
\end{equation}
\ni and
\begin{equation}
\tau_i\nabla_i=\frac{\partial}{\partial \tau}.
\end{equation}
\ni Finally, the normal derivatives of the unit vectors vanish.  The tangential derivatives of
the vectors, which are generally propotional to the curvature by the Frenet relations, are
also taken as zero, since we are ignoring shock curvature.

The derivation of the kinematic shock-generated vorticity proceeds in three stages.  First, we
derive the general expression for the vorticity in normal-tangent coordinates; it is
proportional to the difference between the normal derivative of the tangential velocity
component, $u_{\tau,n}$, and the tangential derivative of the normal velocity component
$u_{n,\tau}$.  We then derive the shock jump conditions for this difference.  The jump
condition for the tangential component of the momentum equation yields an expression for the
jump in the tangential derivative of the pressure $[p_{,\tau}]$ proportional to
$[u_{\tau,n}]$, where brackets refer to changes across the shock.  Taking the tangential
derivative of the normal momentum jump condition gives another expression for
$[p,_\tau]$ which is in this case proportional to $[u_{n,\tau}]$.  Equating the two
expressions for $[p,_\tau]$ results, after some manipulation, in the desired quantity
$[u_{\tau,n}]-[u_{n,\tau}]$, and hence the vorticity jump.  We restrict the analysis to two
dimensions, partly for simplicity, and partly because we were interested in comparisons with
two-dimensional simulations, to be presented elsewhere.  Three-dimensional calculations of
vorticity production by curved shocks in a medium without density fluctuations have been given
by Hayes (1957) and Ram and Upadhyaya (1968).

\subsection {Vorticity in the normal-tangent coordinate system}

The velocity in the normal-tangent coordinate system is $u_i=u_nn_i+u_\tau\tau_i$.  Then the
vorticity is
\begin{eqnarray}
\nonumber\omega_i&=&\epsilon_{ijk}\nabla_ju_k=\epsilon_{ijk}\nabla_j(u_nn_k+u_\tau\tau_k) \\
\nonumber&=&u_n\epsilon_{ijk}n_{k,j}-\epsilon_{ijk}n_ju_{n,k}+\epsilon_{ijk}\nabla_j(u_\tau\tau_k)
\\ 
&=&\epsilon_{ijk}\nabla_j(u_\tau\tau_k)-\epsilon_{ijk}n_ju_{n,\tau}\tau_k.
\end{eqnarray}
\ni The last term involves only the tangential gradient because the normal gradient must
vanish when crossed with the normal vector parallel to it.

In order to express the first term on the rhs of eq. 6 in a useful form, we use the
orthogonality of the unit vectors to obtain
\begin{equation}
0=\nabla_i(u_\tau\tau_jn_j)=\epsilon_{ijk}n_j\epsilon_{klm}\nabla_l(u_\tau\tau_m)+
\epsilon_{ijk}
u_\tau\tau_j\epsilon_{klm}n_{m,l}+u_\tau\tau_j\nabla_jn_i+n_j\nabla_j(u_\tau\tau_i).
\end{equation}
\ni The second term on the right-hand side is zero because it contains a curl of a unit
vector.  The third and fourth terms are directional derivatives.  Taking the first term on
the left-hand side, and using the fact that $n_{i,\tau}=0$ for zero curvature shock surface,
we find
\begin{equation}
\epsilon_{ijk}\epsilon_{jlm}n_k\nabla_l(u_\tau\tau_m)=u_\tau
n_{i,\tau}+\nabla_n(u_\tau\tau_i)= u_{\tau,n}\tau_i.
\end{equation}
\ni Since in two dimensions the vorticity does not have a normal component, equation (1) may
be used to obtain
\begin{equation}
\epsilon_{ijk}\nabla_j(u_\tau\tau_k)=\epsilon_{ijk}n_j\epsilon_{klm}n_m\epsilon_{lqr}\nabla
_q(u_\tau\tau_\tau)=\epsilon_{ijk}n_j(u_{\tau,n}-\kappa u_\tau)\tau_k.
\end{equation}
Substituting (9) into (6) gives
\begin{equation}
\omega_i=\epsilon_{ijk}n_j(u_{\tau,n}-u_{n,\tau})\tau_k.
\end{equation}

\subsection{The Fluid Equations in the normal-Tangent Coordinate System}

The equations of hydrodynamics in cartesian coordinates are the continuity equation
\begin{displaymath}
\rho_{,t}+u_i\rho_{,i}+\rho u_{i,i}=0\ ,
\end{displaymath}
\ni the momentum equation,
\begin{displaymath}
\rho u_{i,t}+\rho u_ju_{i,j}=-p_{,i}
\end{displaymath}
\ni and the energy equation,
\begin{displaymath}
p_{,t}+u_ip_{,i}+\gamma pu_{i,i}=0,
\end{displaymath}
\ni where $\rho$ is the density, $u_i$ is the ith component of the velocity in a cartesian
coordinate system with $i$ ranging from 1 to 3, $p$ is the pressure, and $\gamma$ is the usual
ratio of specific heats.

Expressed in normal-tangent coordinates, the continuity equation becomes
\begin{eqnarray}
\nonumber 0 &=& \rho_{,t}+\nabla_i(\rho(u_nn_i+u_\tau\tau_i)) \\
\nonumber &=&
\rho_{,t}+u_nn_i\nabla_i\rho+u_\tau\tau_i\nabla_i\rho+\rho(\nabla_i(u_nn_i)+\nabla_i(u_\tau\tau_i))
\\
&=& \rho_{,t}+u_n\rho_{,n}+u_\tau\rho_{,\tau}\rho u_{n,n}\rho u_{\tau,\tau},
\end{eqnarray}
\ni where the vanishing of the divergence of the unit vectors was employed.

The momentum equation becomes
\begin{equation}
-\frac{1}{\rho}p_{,i}=u_{n,t}n_i+u_{r,t}\tau_i+(u_nn_j+u_\tau\tau_j)\nabla_j(u_nn_i+u_\tau\tau_i).
\end{equation}
\ni The normal derivatives are
\begin{equation}
u_nn_j\nabla_j(u_nn_i+u_\tau\tau_i)=u_nu_{n,n}n_i+u_nu_{\tau,n}\tau_i.
\end{equation}
\ni The tangential derivatives are
\begin{equation}
u_\tau\tau_j\nabla_j(u_nn_i+u_\tau\tau_i)=u_\tau u_{n,\tau}n_i+u_\tau u_{\tau,\tau}\tau_i.
\end{equation}

The above equations can be combined to give the normal
\begin{equation}
\rho(u_{n,t}+u_nu_{n,n}+u_\tau u_{n,\tau})=-p_{,n}
\end{equation}
\ni and tangential
\begin{equation}
\rho(u_{\tau,t}+u_nu_{\tau,n}+u_\tau u_{\tau,\tau})=-p_{,\tau}
\end{equation}
\ni components of the momentum equation.

The energy equation becomes
\begin{eqnarray}
\nonumber 0&=&p_{,t}+(u_nn_i+u_\tau\tau_i)\nabla_ip+\gamma p\nabla_i(u_nn_i+u_\tau\tau_i) \\
&=& p_{,t}+u_np_{,n}+u_\tau p_{,\tau}+\gamma p(u_{n,n}+u_{\tau,\tau}).
\end{eqnarray}

\subsection {Oblique Shock Jump Conditions}

Let the quantities in front of the shock be labeled with subscript 0.  These are the
preshock density $\rho_0$, pressure $p_o$, and velocity $u_o$.  The respective postshock
quantities ($\rho,\ p,\ u_i$) will have subscript 1.  For the shock jump conditions, we
introduce the notation $[a]\equiv a_1-a_0$.  Then the shock jump conditions are mass
conservation
\begin{equation}
[\rho u_n]=0,
\end{equation}
\ni normal momentum conservation
\begin{equation}
[p+\rho u^2_n]=0,
\end{equation}
\ni tangential momentum conservation
\begin{equation}
[u_\tau]=0\ ,
\end{equation}
\ni and energy conservation
\begin{equation}
\left[\frac{1}{2}u^2_n+\frac{\gamma}{\gamma-1}\frac{p}{\rho}\right]=0.
\end{equation}
\ni By equation (18)
\begin{equation}
[u_n]=(\rho_0u_{0n}/\rho)-u_{0n}=u_{0n}(\rho_0-\rho_1)/\rho_1.
\end{equation}
\ni For determining the change in vorticity it is convenient to introduce the relative density
jump,
\begin{equation}
\Delta\equiv(\rho_1-\rho_0)/\rho_0.
\end{equation}
\ni Then $\rho=(1+\Delta)\rho_0$ and $\rho_0-\rho_1=-\Delta\rho_0$, so
\begin{equation}
[\rho]=\Delta\rho_0
\end{equation}
\ni and
\begin{equation}
[u_n]=-u_{0n}\Delta /(1+\Delta).
\end{equation}
\ni Using (18) and (19), the pressure jump is
\begin{equation}
[p]=-[\rho u^2_n]=-\rho_0u_{0n}[u_n].
\end{equation}
Let $c_s\equiv\sqrt{\gamma p_0/\rho_0}$ be the sound speed.  Then the Mach number is $M\equiv
u/c_s$.  However, for an oblique shock, it is convenient to introduce the quantity $M_n\equiv
u_n/c_s$.  Then combining equations (18--21) gives the solutions
\begin{equation}
\rho_1=\frac{(\gamma+1)M^2_n}{(\gamma-1)M^2_n+2}\rho_0 
\label{eq:32a}
\end{equation}
\ni and
\begin{equation}
p_1=\frac{1+\gamma(2M^2_n-1)}{\gamma+1}p_0
\label{eq:32b}
\end{equation}

\subsection {Vorticity Jump Condition}

The jump condition for the tangential component of the momentum equation (16) is found by
substituting eqs. 18 and 20, and using the fact that $u_{\tau,\tau}=0$ for non-curved shocks. 
The result is
\begin{equation}
-[p_{,\tau}]=\rho_0u_{0n}[u_{\tau,n}]+[\rho]u_{0\tau}u_{0\tau,\tau}=\rho_0u_{0n}[u_{\tau,n}].
\end{equation}
\ni Next, the tangential derivative of the normal momentum jump condition (19) gives, using
eq. (18) 
\begin{equation}
-[p_{,\tau}]=\delta_\tau u_{0n}[u_n]+\rho_0u_{0n,\tau}[u_n]+\rho_0u_{0n}[u_{n,\tau}].
\end{equation}
\ni Finally, we need an expression for $u_{0n,\tau}$, which we next derive for isobaric preshock density
fluctuations, under the assumption that the postshock pressure is constant along the shock.  This assumption
depends on, among other things, considerations of the effect of a reflected shock on the postshock
environment.  An estimate for the magnitude of this effect on the constant postshock pressure assumption is
attempted in the Appendix.

We now turn to the derivation of $u_{0n,\tau}$, which occurs in eq. (30).  Because the shock
jump conditions involve the quantities where the shock is at rest, we consider infinitesimal
regions of the shock line such that a transformation may be made to a coordinate system such
that adjacent regions along the shock line are also at rest.  Since the preshock velocity is assumed to be
zero, at any point in this frame $u_{0\tau}=0$, and the normal component $u_{0n}=-v_{sh}=-Mc_0$.  In
this case
\begin{eqnarray}
\nonumber u_{0n,\tau}&=& -\frac{\partial v_{sh}}{\partial\tau} \\
\nonumber &=& -\frac{\partial}{\partial\tau}\left(M\sqrt{\gamma p_0/\rho_0}\right) \\
\nonumber &=& -M_{,\tau}\sqrt{\frac{\gamma
p_0}{\rho_0}}-\frac{Mp_{0,\tau}}{2}\sqrt{\frac{\gamma}{p_0\rho_0}}-M\sqrt{\gamma
p_0}\left(-\frac{1}{2\rho^{3/2}_0}\delta_\tau\right) \\
\nonumber &=&-\frac{M_{,\tau}v_{sh}}{M}-\frac{v_{sh}p_{0,\tau}}{2p_0}+\frac{v_{sh}
\delta_\tau}{2\rho_0} \\ 
&=&\frac{M_{,\tau}u_{0n}}{M}+\frac{u_{0n}p_{0,\tau}}{2p_0}-\frac{u_{0n}\delta_\tau}{2\rho_0}
\end{eqnarray}
\ni In the above expressions we denote the tangential derivative of the preshock density by
\begin{equation}
\delta_\tau\equiv\partial\rho_0/\partial\tau
\end{equation}
\ni In this frame, because the tangential gradient of the normal velocity does not vanish, it
may appear that there is some pre-existing vorticity in the preschock region.  However, this
is not the case.  Because the different parts of the shock line are not traveling at uniform
speed, a transformation to a frame in which the shock line is at rest is not a Galilean
transformation.  This is why infinitesimal regions must be considered.

In the isobaric case, the second term in equation (31) vanishes by definition.  Also, since
the pressure in front of the shock is constant (by the isobaric assumption) and the pressure
behind the shock is constant (by the main assumption of this section), the pressure jump is
constant along the shock.  By the equation for the pressure jump across the shock (28), the
only way that can occur is if the Mach number $M=$ constant.  (It is important to notice that
this is the only point in the derivation where the energy equation is used.)  Therefore, the
first term also vanishes.

Then equation (31) reduces to
\begin{equation}
u_{0n,\tau}=-\frac{u_{0n}\delta_\tau}{2\rho_0}
\end{equation}
Substituting eq. (33) for $u_{0n,\tau}$ in eq. (30) gives 
\begin{equation}
-[p_{,\tau}]=\delta_\tau u_{0n}[u_n]+\rho_0u_{0n}[u_{n,\tau}]
\end{equation}
\ni where $\delta_\tau=\partial\rho_0/\partial\tau$.  Equating the two expressions (34) and
(29) and using the jump condition $[u_n]=-u_{0n}\Delta/(1+\Delta$), where
$\Delta=(\rho_1-\rho_0)/\rho_0$ is the fractional density jump, gives
\begin{equation}
[u_{\tau,n}]-[u_{n,\tau}]=\frac{\delta_\tau}{\rho_0}\frac{u_{0n}\Delta}{(1+\Delta)}
\end{equation}
\ni Substitution into eq. (10) for the tangential vorticity component, assuming no vorticity
in the preshock cloud, and using $u_{0n}=-v_{sh}$ (in the shock frame) finally gives the
generated kinematic vorticity just behind the shock at any position ($r,\theta$) in the cloud
as
\begin{equation}
\omega(r,\theta)=v_{sh}(r)\delta_\tau(r,\theta)\Delta/\rho_0(r)(1+\Delta).
\end{equation}
\ni Eq. (36) is the main result of this section.  For an isobaric preshock density distribution this
expression gives the kinematically generated vorticity as a function of position due to the
passage of the shock.  It should be noted that we have neglected the compression of the cloud by the shock
passage, which is why the radial coordinates on the lhs and rhs of eq. (36) refer to the same position.
%\clearpage

\section{SHOCK INTERACTION WITH A RADIAL DENSITY GRADIENT}
\subsection{Spatial dependence of kinematic shock-generated vorticity}
We evaluate the shock-induced vorticity as a function of position in the cloud for the
following density distribution:
\begin{equation}
\rho_o(r)=\rho_a+\frac{\rho_c-\rho_a}{1+(r/R_o)^n}=\rho_a\left(\frac{\xi+\chi-1}{\xi}\right)
\end{equation}
\ni where $\xi(r)\equiv1+(r/R_o)^n$, $R_o$ is a characteristic radius, and the density
contrast is $\chi\equiv\rho_c/\rho_a$.  We have also considered an exponential density
distribution, but the algebra is sufficiently complicated, and the results so similar, that we
only indicate the differences when appropriate.  

For the power law density distribution the
tangential density derivative is (neglecting the curvature)
\begin{equation}
\delta_\tau(r,\theta)=\frac{\partial\rho_a}{\partial y}=-\rho_a(\chi-1)nr^{n-1}\
{\rm sin}\theta/R_o^n\xi^2
\end{equation}
The shock speed $v_{sh}$ in the cloud is a function of radius, and can be expressed in terms
of the incident shock speed $v_\infty\left(={\mit M}\sqrt\frac{\gamma p_o}{\rho_a}\ \right)$
by
\begin{equation}
v_{sh}(r)=v_\infty[(\xi/(\xi+\chi-1)]^{1/2}\ .
\end{equation}
Also, notice that for the isobaric clouds considered here, the pressure difference across
the shock interface is constant along the shock surface, so the Mach number and $\Delta$ are
constant.  Combining these factors in eq. (36) gives
\begin{equation}
\omega(r,\theta)=-\frac{nv_\infty\Delta(\chi-1)}{2(1+\Delta)R_o^n}\ \frac{r^{n-1}\
{\rm sin}\theta}{(\xi+\chi-1)^{3/2}\xi^{1/2}}
\end{equation}
Because of the sin$\theta$ factor, the vorticity will be concentrated in regions above and
below (along the y-axis) of the center of the cloud.  The dependence on $\xi$ tends to make
these regions flattened along the y-axis.  The ratio of specific heats $\gamma$ and the
Mach number of the incident shock $M=v_{sh}/c_s$ (= constant) enter through the
fractional density jump given by $1+\Delta=(\gamma+1)M^2/[(\gamma-1)M^2+2]$,
which for a strong shock is independent of $M$ and equals 4 for $\gamma=5/3$.

Figure 2 shows a grey-scale representation of the induced vorticity given by eq. (40) (or its
equivalent for the exponential density distribution) for four forms of the density
distribution, with central density enhancement $\chi=10$.  Figure 3 shows the angle-averaged
radial distribution of vorticity (in units of the shock velocity divided by $R_o$) for
the same four density distribution, each for four values of $\chi$.  It can be seen that
increasing the density contrast $\chi$ causes the position of peak induced vorticity to
shift further away from the center of the cloud relative to the scale radius $R_o$.  The
important point is that, unless the density gradient is very steep or the density contrast
very large, much of the vorticity is generated well inside the cloud.

\begin{figure}
\plotone{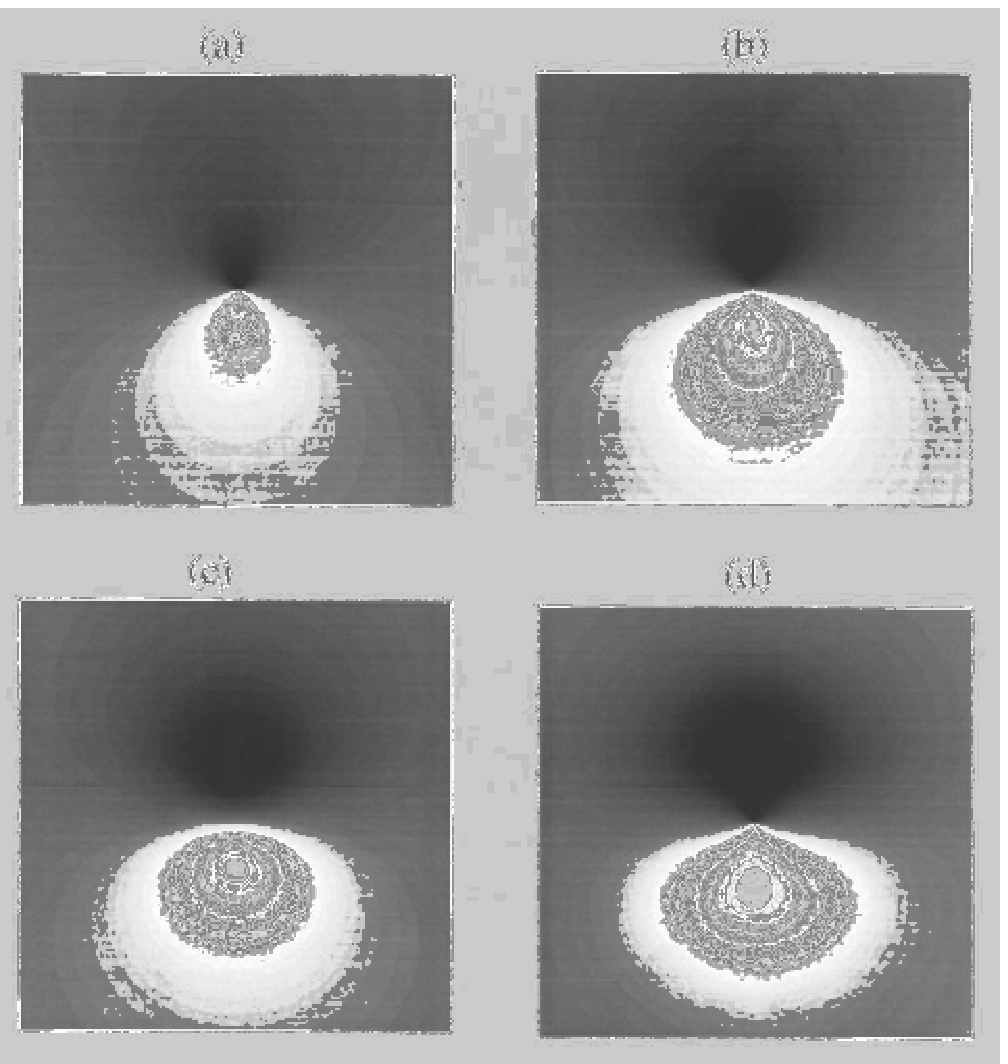}
\figcaption{Greyscale map of the induced kinematic vorticity within a model cloud for power
law density distributions with $n=1$ (a), $n=3/2$ (b), $n=2$ (c), and for an exponential
density distribution (d).  The density contrast is $\chi=10$ and each map extends to 5 scale
radii.}
\end{figure}

\begin{figure}
\plotone{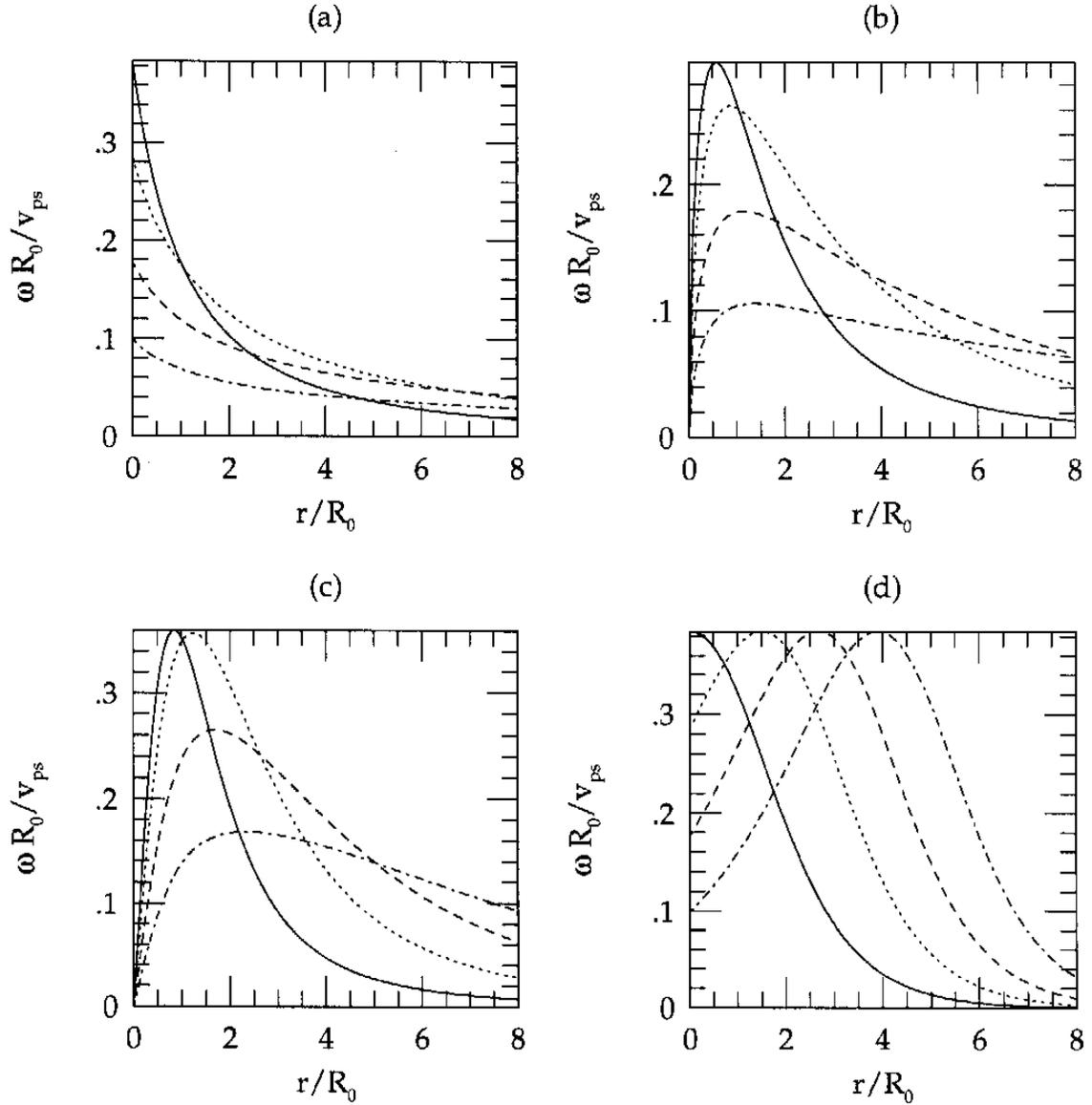}
\figcaption{Angle-averaged radial dependence of shock-induced kinematic vorticity, in units
of postshock speed divided by scale radius, for the four density distributions (three power
laws and an exponential) used in Fig. 2.  The solid lines correspond to density contrast
$\chi=3$, dotted lines to $\chi=10$, dashed lines to $\chi=30$, and dashed-dotted lines to
$\chi=100$.}
\end{figure}

\subsection{rms kinematic vorticity}
In order to characterize the magnitude of the kinematically-generated vorticity due to the
shock passage, we calculate the rms vorticity in a region of size $R$ and mass $m$:
\begin{equation}
\omega_{rms}\equiv\left[\frac{1}{m}\int\rho\omega^2dA\right]^{1/2}=\left(       
\frac{1}{m}\int^R_0\int^{2\pi}_0\
\rho(r)\omega^2(r,\theta)rdrd\theta\right)^{1/2}
\end{equation}

Letting $z\equiv R/R_o$, we find
\begin{equation}
\int\rho \omega^2dA=\frac{\pi\rho_a\Delta^2v^2_\infty
n}{4(\Delta+1)^2}\times\left[\frac{(\chi+1)\ell
n\left(\frac{(1+z^n)\chi}{\chi+z^n}\right)}{\chi-1}+\frac{1}{1+z^n}+\frac{\chi}{\chi+z^n}-
2\right]
\end{equation}

The mass m contained in a region of radius $R$ is given by $m=2\pi\int^R_0\rho(r)rdr$.  For
the density distribution given by eq. (37), the integral can be obtained in closed form for
only selected values of the exponent $n$. For the cases
$n=1$, 3/2, and 2 we find
\begin{equation}
m= \rho_a\pi R_o^2[z^2+b(\chi-1)]
\end{equation}
\ni where $b=2[z-\ell n(1+z)]$ for $n=1$ and $b=\ell n(1+z^2)$ for $n=2$.  For $\chi\ll
z^2$, the first term dominates and $m$ is independent of the density contrast $\chi$,
while in the limit of large $\chi,\ m\propto\chi$, as expected.  For $n=3/2$ and for
an exponential density distribution the expression for $b$ is more complicated, but m
has the same limits, and we assume that the small and large $\chi$ results are
general.
 
In the limit $\chi\rightarrow1(\chi\ll z^2)$, the integral (42) can be shown to be
proportional to $\chi-1$ and m (eq. 43) is independent of $\chi$, so
$w_{rms}\propto(\chi-1)^{1/2}$.  In the large $\chi$ limit, the integral (42) is
\begin{equation}
\int\rho\omega^2dA=\frac{\pi\rho_a\Delta^2v^2_\infty
n}{4(\Delta+1)^2}\times\left[\ell n(1+z^n)+\frac{1}{1+z^n}-1\right]^{1/2}
\end{equation}
\ni independent of $\chi$.  Since the mass is proportional to $\chi$ for $\chi\gg z^2$,
$\omega_{rms}\propto\chi^{-1/2}$ in the large $\chi$ limit.

For an exponential density distribution the algebra is more complicated, but the same results
for the $\chi$-dependence are found in the small- and large-$\chi$ limits.

The result that $\omega_{rms}$ scales with density contrast as $\chi^{-1/2}$ in the limit of
large density contrast is different than found by Klein et al. (1994) for a sharp-boundaried
cloud, who found that the average vorticity is {\it independent} of $\chi$ in the large $\chi$
limit.  The reason for this difference is that for a sharp-boundaried cloud the vorticity is
confined to a small boundary layer, and so depends on the {\it difference} between the
postshock speeds inside the cloud and in the ambient medium.  This difference, which is
proportional to ($1-\chi^{-1/2}$), is constant and equal to 1 for large $\chi$.  For clouds
with a continuous density gradient the vorticity is located inside the cloud and thus just
depends on the postshock speed in the cloud.  As discussed above, the average postshock speed
inside the cloud is proportional to $\chi^{-1/2}$.

It is useful to re-express the above results in terms of the quantity
$\beta_{kin}=w_{rms}/(v_\infty/R)$, which may be considered as the ratio of the characterstic
vortical velocity $v_{vor}\equiv w_{rms}R$ to the shock speed in the ambient medium
$v_\infty$.  In the limit of large $\chi$ and $z>1$ we find
\begin{equation}
\beta_{kin}=\frac{\Delta zn^{1/2}}{2(\Delta+1)\chi^{1/2}}\left(\frac{\ell
n(1+z^n)}{f(z)}\right)^{1/2}
\end{equation}
\ni where $f(z)=2z$, 4z$^{1/2}$, and $\ell n(1+z^2)$ for $n=1$, 3/2, and 2, respectively. 
Assuming $z\sim1$,
\begin{equation}
v_{vor}/v_\infty\approx\frac{\Delta n^{1/2}}{2(\Delta+1)}\chi^{-1/2}
\end{equation}
For the exponential density distribution, the coefficient is $\Delta/[2\sqrt{2}(\Delta+1)]$. 
In the limit of large Mach number, $\Delta/(\Delta+1)=3/4$ for $\gamma=5/3$, or $\sim1$ for
$\gamma=1$.  So for large $\chi$ and $M$, a fraction
$3\sqrt{n}\chi^{-1/2}/8\ (3\chi^{-1/2}/8\sqrt{2}$ for an exponential distribution) of the
ambient shock speed may be transmitted into a cloud as a vortical mode.

The amplitude and dependence on Mach number and density contrast predicted by eq. 46 have been checked
by comparison with lattice gas hydrodynamical simulations of shock-cloud interactions in two
dimensions (Kornreich \& Scalo 2000).  Because of the nature of the lattice-gas method, only
simulations with Mach numbers up to about 3 were possible.  However, within this constraint, the
simulations (which include the flattening of the cloud and the shock curvature) show good agreement
with eq. 46.  In addition, the simulations verify that most of the vorticity is generated as two
oppositely-signed vortices {\it within} the cloud (rather than a vorticity sheet at the surface), and
that there is little tendency for cometary morphology after shock passage, or shredding from the outer
layers of the cloud, as has been found in previous simulations of sharp-edged clouds (see sec. 3.6.
below), at least for the modest Mach numbers accessible to the simulations.  The simulations also verify that
the dynamical vorticity generation (as opposed to the kinematic vorticity generation studied here) is
negligible except at Mach numbers close to unity.  

\subsection{Density scaling}
Figure 4 shows $\beta_{kin}$, calculated using eqs. 43--45 (or, for eqs. 42 and 44, from the
corresponding equations for the exponential case), as a function of $\chi$, for four
different density distributions and four Mach numbers.  The adiabatic index was taken to be
$\gamma=5/3$.  Changing the adiabatic index to $\gamma=1$ only increases $\beta_{kin}$
slightly, since then, for example, $\Delta/(\Delta+1)\approx1$ instead of 3/4 at large Mach
numbers.  The radial limit of integration was taken as $R/R_o=2$.  The shape of the curves
does not depend on the adopted density distribution; only the maximum value of $\beta_{kin}$
and the value of
$\chi$ for which it occurs depend on the density distribution, and only weakly.  The plots
illustrate the limiting forms given earlier.  However notice that the $\chi^{-1/2}$
dependence is only valid for density contrasts greater than about 6--10.  At small density
contrast the curves increase with $\chi$, roughly in agreement with the ($\chi-1)^{1/2}$
dependence (which is a steep function for $\chi\approx1$) predicted above.

\begin{figure}
\plotone{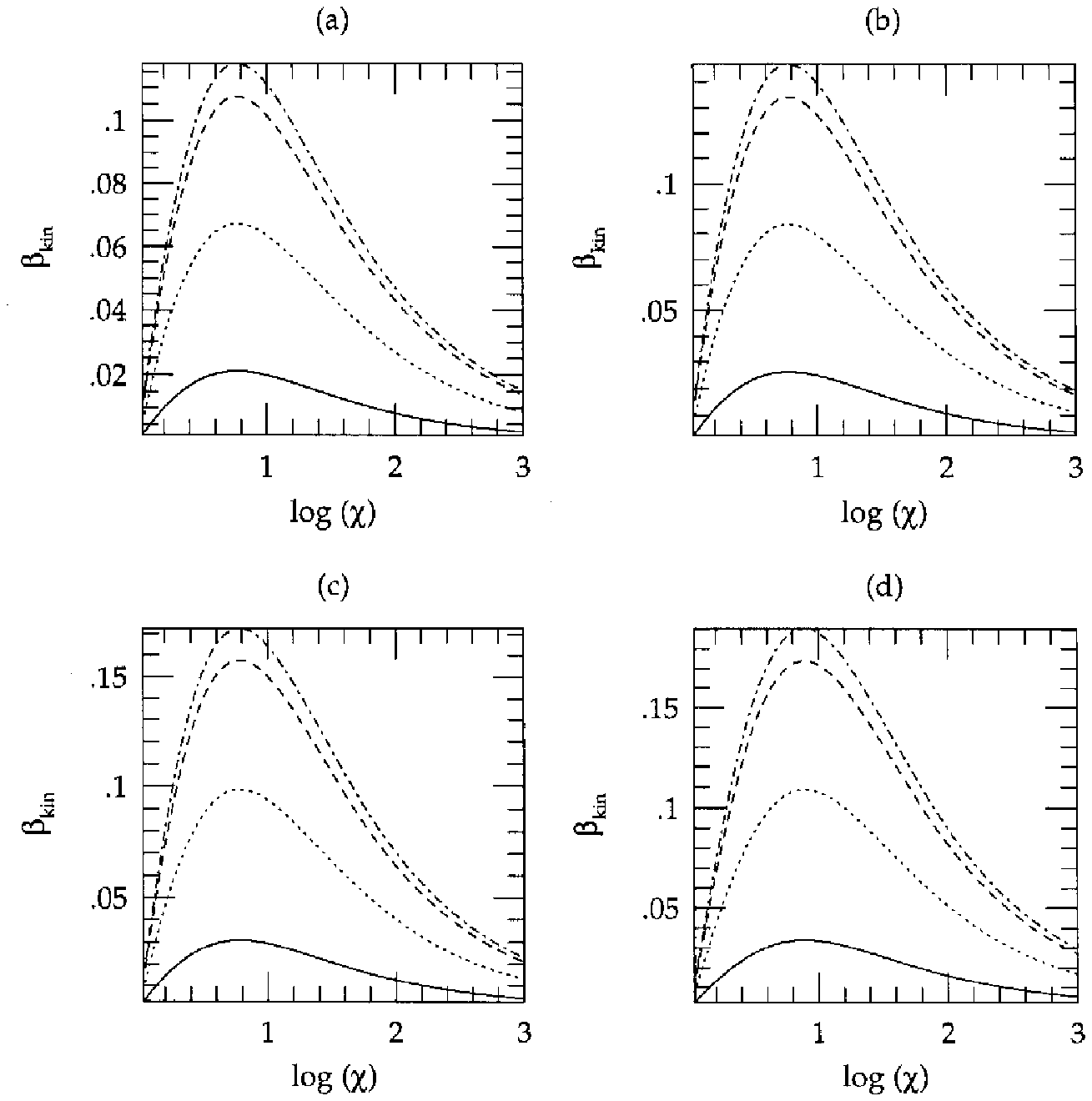}
\figcaption{Ratio of the characteristic induced kinematic vortical speed $\omega_{rms}$R to
the shock speed in the ambient medium as a function of the logarithm of the density
contrast $\chi$ for $\gamma=5/3$ and a radial integration limit $R/R_o=2$.  The four panels
correspond to power law density distributions with indices 1 (a), 3/2 (b), 2 (c) and an
exponential distribution (d).  In each panel the curves correspond to Mach numbers of 1.1
(solid), 1.5 (dotted), 3 (dashed), and 6 (dashed-dotted).}
\end{figure}

The scaling predicted by Chernin and collaborators (Chernin 1996 and references therein),
which is based on a scattering approach that assumes small $\chi$ and large Mach number,
also increases with $\chi$.  They give the dependence of the spatially maximum value of
$\beta$ as proportional to $\chi$, while we derive the rms value of $\beta$, so it isn't
clear whether there is a discrepancy.  In any case, we have not assumed either small density
contrasts or large Mach numbers in the above derivation, and our small-$\chi$ case is just a
limit of the general case, while Chernin et al. assume both limits at the outset.  Given
that the $\chi$-dependence changes its functional form at larger density contrast, it is
inappropriate to apply the small-$\chi$ case to cloud complexes, GMCs, etc. as do Chernin and
Efremov (1995).  However it is still important to note that the spatial distribution of
vorticity found in their quasi-hydrodynamical simulations, which still use the assumptions
of small density contrast and large Mach number to calculate velocity perturbations behind
the shock, is similar to that found here, namely two counter-rotating vortices.  Chernin
and Efremov also illustrate interesting cases of a shock passing obliquely through an
elongated two-dimensional cloud, and a three-dimensional calculation.

\subsection{Validity of the frozen vorticity assumption}
The above derivation assumes that the induced vorticity does not change during the shock
passage through the cloud.  The viscous decay time is of order $\tau_{vis}\sim \ell^2/\lambda
v_{th}$ where $\ell$ is the scale on which the vorticity is generated, $\lambda$ is the mean
free path, and
$v_{th}$ is the thermal speed.  Taking
$\lambda=1/n\sigma$, where
$n$ is the number density and $\sigma$ is the particle cross section, and assuming
$\sigma\sim10^{-15}cm^2$ and $n\sim10^3R_{pc}^{-1}cm^{-3}$ for illustration, we find 
\begin{equation}
\tau_{vis}\sim3\times10^6\ell^2/Rv_{th} ,
\end{equation}
\ni which is much larger than the timescale for shock passage, $\tau_{sh}\equiv R/v_{sh}$
for any reasonable value of $\ell$.  However the induced vorticity will also change due to
conversion into other fluid modes (dilatation, MHD modes; see below).  The timescale for these
changes is expected to be of order
$\tau_{vor}\sim R/v_{vor}$.  Then $\tau_{vor}/\tau_{sh}=2[(\Delta+1)/\Delta]\ell/R$ which, for
large Mach number, is (8/3)$\ell$/R for $\gamma=5/3$ and 2 $\ell$/R for $\gamma=1$.  So while
the vorticity is not strictly ``frozen," the vorticity should not change appreciably during
shock passage.  The frozen vorticity assumption becomes better for weaker shocks.

\subsection{Compressional and bulk components}
For the conditions studied here, the local induced motions within the cloud should
initially be composed equally of compressional and vortical modes.  The conditions that
provide this equivalence are the radially symmetric density gradients and the nonoblique
shock.  Since the density gradient is radially symmetric, the density gradients parallel and
perpendicular to the shock propagation direction are equal.  Taking into account the
assumption about the constant pressure behind (and in front of) the shock, by momentum
conservation across the shock, the postshock velocity gradients are determined by the
preshock density gradients, and therefore should also be symmetric (ignoring such distortions
as cloud flattening).  Therefore, since the vorticity may be thought of as the gradient in
the velocity perpendicular to the direction of the shock propagation, and the dilatation is
due to gradients in the velocity parallel to the direction of the shock propagation, the
vorticity and the dilatation should be comparable.  The nonoblique condition means that the
cloud had no (initial) tangential velocity relative to the shock.  Thus, the postshock
velocity is dominated by motions in the direction of the shock (i.e. the component of the
velocity perpendicular to the shock propagation, which would be conserved across the shock, is
small).  It follows from these two conditions that since the only source of the velocity
behind the shock is the momentum imparted from the shock, and the gradients of the postshock
velocity parallel and perpendicular to the shock propagation are equivalent, then the
vortical and compressional components of the velocity must be (approximately) equal. 
In a statistical sense, the same conclusion may be drawn for any inhomogeneous density
medium that does not have a preferential gradient; i.e. the {\it average} compressional and
vortical velocities should be comparable.  An example of the violation of the
nonpreferential density gradient would be if, say, all the clouds were prolate with their
long axes in the direction of the shock.  The reason that this effect (i.e. the equivalence
of the vortical and compressional velocities) would be violated for an oblique shock is the
differences in the way the curl and the divergence are taken and, in particular, the manner
in which the gradients of the tangential component of the velocity are affected by the
shock.  Because of the curvature of the shock near the center of the cloud, oblique shock
effects will also occur even in the case of symmetric preshock gradient, and therefore any
distortions in the regions where the shock is significantly curved will not be transitory. 
Still, to first order it is expected that the compressional and vortical velocities, at
least initially, should be comparable.

Besides the internal vortical and compressional modes excited by the shock, the shock also
accelerates the whole cloud.  The bulk cloud speed is the average postshock speed $v_{ps}$
in the cloud:
\begin{equation}
v_{cl}=\int\rho v_{ps}(r)/m
\end{equation}
The postshock speed is $v_{ps}=\frac{\Delta}{\Delta+1}v_{sh}(r)$, where $v_{sh}(r)$ is
given by eq. (41).  Taking $\Delta/(\Delta+1)$ out of the integral because of the
constant Mach number assumption, we find, for large $\chi$,
\begin{equation}
v_{cl}=\frac{\Delta}{\Delta+1}\chi^{-1/2}v_\infty\ .
\end{equation}

This result is actually the same as that for a cloud with a sharp boundary (McKee \& Cowie
1975, Miesch \& Zweibel 1994).  The reason for the equivalence is that in the limit of large
$\chi$, the mass in the continuous case is more concentrated in the center of the cloud, and
thus the mass weighted bulk motion approaches the velocity at the center of the cloud.  Using
the induced vortical velocity from eq. (48), taking $n\approx1$, this gives
$v_{cl}\approx2v_{vor}$.  By the argument of the previous paragraphs the vortical velocity is
about equal to the compressional velocity $v_{comp}$, so
$v_{cl}\approx v_{vor}+v_{com}$, and there is approximate equipartition between the bulk
cloud speed and the total induced internal motions.  Notice that the definition of the
``large $\chi$ limit" ($\chi\gg(R/R_o)^n$) depends on the limit of integration 
and $n$, since for larger R the now fast-moving material is averaged with the cloud speed,
but the gradient in the velocity is smaller; also, in the central regions the shock curvature
will be significant.  However, since the location of maximum vorticity is away from the
center, this result should be valid to order of magnitude.

\subsection{Comparison of induced velocity amplitudes with observations}
Both the internal and bulk components of the shock-induced velocity scale with cloud
density as $\rho^{-1/2}$.  Although we do not wish to place too much emphasis on this result
until the statistical correlations for an {\it ensemble} of shock-cloud interactions is
calculated, it may be significant that the ``standard" empirical scaling relations give this
same linewidth-density scaling if size is eliminated, and it may be, as suggested by Xie
(1997) for a very different model, that it is the linewidth-density scaling that is most
physically fundamental.  {\it If}
$\rho\propto R^{-1}$, as is often claimed, the induced motions, whether for the internal
motion or for an ensemble of clouds, would satisfy a linewidth-size relation $\Delta v\propto
R^{1/2}$.  We are unable to justify such a density-size relation, since we cannot
derive the characteristic sizes of the induced velocity and density fluctuations
analytically.  However the $\Delta v\propto R^{1/2}$ relation in the sense of a scaling of
the velocity structure function at least will automatically be satisfied by an
uncorrelated field of discontinuities (Saffman 1971), although modifications will
certainly appear in more realistic situations.  In this case the shock pump could in
principle account for both observed (uncertain) scaling relations, without requiring
self-gravity, magnetic fields, or internal protostars.  

More importantly for the present work, the shock-induced
velocities are consistent with the {\it amplitudes} of the observed linewidths.  If we
identify the shock with supernova remnants or superbubbles, the most frequent shock speed by
far will be the terminal speed (because the distribution of shell speeds is a strongly
decreasing function of speeds, see sec. 4 below), which is usually taken to be $\sim10km\
s^{-1}$.  Assuming $\rho\sim R^{-1}$ for illustration, on the large end of the spectrum of
GMC sizes, 100 pc, we find $\Delta
v\sim\frac{\Delta}{2(\Delta+1)}R^{1/2}\sim(3-8)R^{1/2}_{100pc}\ km\ s^{-1}$. If smaller
clouds are nested within these larger structures, then the transmitted shock from the shock
passage through the larger cloud is decreased by a factor of
$\chi^{-1/2}$, giving the correct amplitude for the smaller scales.  Of course this argument
is too crude, since shocks will be generated within clouds of various densities.  A treatment
of the relation between the statistical properties of the density and velocity fluctuations is
postponed to future work.

There is an important difference between the morphology of shocked clouds predicted by the present
model and that predicted by previous simulations of shocks incident on clouds with sharp boundaries. 
In the latter case the vorticity is generated in a sheet at the model cloud surface, and this material
is swept behind the cloud by the shock, resulting in comet-shaped clouds (or filaments with
``heads").  Such structures are found even in simulations of low-velocity shock-cloud interactions
(Horvath \& Toth 1995).  However the point of the present work is that, when the cloud possesses a
continuous density gradient, most of the vorticity is generated {\it within} the cloud, not at its
boundary, and thus there is much less tendency for the shock to ``shear off" the edge of the cloud. 
Two-dimensional lattice gas simulations of shock-cloud interactions up to Mach numbers of 3 (KS00)
confirm this picture: the simulated clouds flatten during the shock passage, but most of the
vorticity generation is internal and little of the exterior portions of the cloud are ``shredded."  At
larger Mach numbers the shredding may be more significant because of the increased flattening (or
instabilities, see Schiano et al. 1995), but in general we do not predict the degree of cometary, or
head-tail, structure predicted by previous studies of sharp-edged clouds.  Therefore, although some
cometary structures are observed in the ISM, the fact that most clouds do {\it not} exhibit such
structure may be interpreted as support for our model.    

\subsection{Fate of the induced vorticity}
The kinematic vorticity generated by a shock passing through a density inomogeneity is
unlikely to remain in the vortical mode for long.  In sec. 3.4. we showed that the viscous
decay is negligible.  The vorticity
could be amplified in the postshock flow by the baroclinic vector due to the generally
non-parallel pressure and density gradients in the postshock flow, but an analysis of this
effect (KS00) indicates that the vorticity amplificaton will be negligible
except at Mach numbers close to unity.  In three dimensions, the vorticity can amplify or
suppress itself through the vorticity stretching term in the momentum equation, but there is
good reason to think that the dominant process will be the conversion of vorticity into
compressible modes.

One way to see this is to examine the evolution equations for the vorticity
$\omega=(\nabla\times u$) and the dilatation $\theta\equiv\nabla\cdot u$.  Taking the curl of
the momentum equation, the evolution equation for a component $\omega_i$ of the vorticity can
be shown to be
\begin{equation}
\frac{d\omega_i}{dt}=\omega_j\sigma_{ij}-\frac{2}{3}\theta\omega_i-\frac{\epsilon_{ijk}p_{ij}\rho_{jk}}
{\rho^2}
\end{equation}
\ni where $\sigma_{ij}$ is the symmetric shear tensor given in terms of the velocity gradient
tensor $\psi_{ij}\equiv u_{i,j}$ by 
\begin{equation}
\sigma_{ij}\equiv\frac{1}{2}[\Psi_{ij}+\Psi_{ji}]-\delta_{(ij)}\Psi_{ij}/3\ ,
\end{equation}
\ni where the parentheses on the subscripts of the Kronecker delta indicate that there is no
summation.  The first two terms on the right-hand side of eq. (50) are equivalent to the
vorticity stretching term $\omega\cdot\nabla u$, but we have decomposed the velocity gradient
tensor to make explicit the interaction between vorticity and dilatation.  See Peebles (1993,
sec. 22) for a similar derivation in a cosmological context.  In terms of the positive
quantity $\omega\equiv(\omega_i\omega_i)^{1/2}$ the vorticity equation becomes 
\begin{equation}
\frac{dw}{dt}=\frac{1}{\omega}\omega_i\omega_j\sigma_{ij}-\frac{2}{3}\theta\omega-
\frac{\omega_i\epsilon_{ijk}p_{,j}\rho_{,k}}{\omega\rho^2}\ .
\end{equation}

The divergence of the momentum equation gives
\begin{equation}
\frac{d\theta}{dt}=\frac{1}{2}\omega^2-\sigma^2_{ij}-\frac{1}{3}\theta^2-\frac{1}{\rho}
\nabla^2p+\frac{\rho_{,i}p_{,i}}{\rho^2}\ .
\end{equation}
Ignoring all terms besides the explicit vorticity-dilatation interactions, it can be seen that
there is an asymmetry between the two equations.  The magnitude of the vorticity can only
decrease due to the vorticity-dilatation interaction, but this vorticity always increases the
dilatation.  Another way to see this is to examine the evolution equation for the ratio of
vorticity magnitude to dilatation.  Again neglecting all terms except those explicitly
involving the vorticity and dilatation, the result is
\begin{equation}
\frac{d(\omega/\theta)}{dt}=-\frac{\omega}{3}-\frac{\omega^3}{2\theta^2}
\end{equation}
\ni This asymmetry (in a somewhat different form) was pointed out independently by
Vazquez-Semadeni, Passot, \& Pouquet (1996), who also present evidence from simulations that,
in highly compressible flows, vortex stretching acts primarily to drain the vorticity. 
Additional support for the dominance of transfer from vortical to compressible modes comes
from the eddy-damped quasinormal Markovian closure analysis of weakly compressible turbulence by
Bataille \& Zhou (1999).  The ability of a purely incompressible (vortical) velocity field
to generate density fluctuations is also shown in the 2-dimensional simulations of Chantry,
Grappin, \& Leorat (1993) for a simple initial velocity field.

The upshot is that we expect the shock-generated vorticity to be converted to compressible
modes, and the associated density inhomogeneities (since $d\ell n\rho/dt=-\theta$).  It is important
to realize that this mechanism provides a source of large-amplitude density fluctuations that
do not owe their origin to any sort of instability, gravitational or otherwise.  Since the
induced vortical velocities are generally supersonic for the interstellar conditions
considered here, we expect the induced vorticity to result in a system of internal shock
waves, which themselves will generate their own kinematic vorticity and dilatation on smaller
scales.  Thus {\it shock-generated kinematic vorticity provides a mechanism for top-down
cascade of kinetic energy and creation of a nested system of density fluctuations.}  We
suggest that this process could be at least part of the physics behind the observed
fractal-like structure of the cool interstellar medium, an empirical property that has many
potentially important implications (Elmegreen \& Falgarone 1996, Elmegreen 1997a,b).

There is ample evidence that complex density structure exists in regions in which 
self-gravity is unimportant and there are no stellar power sources.  A good example is the
densely sampled map of the ``Polaris Flare" cloud by Heithausen \& Thaddeus (1990). The study
of the cloud MBM 12 (size $\sim$2 pc) by Pound et al. (1992) reveals numerous
``clumps" with supersonic linewidths and sizes down to the resolution limit of $\sim$0.03
pc.  Similar results have been obtained for the cloud MBM 7 by Minh et al. (1996), who suggest
that the large turbulent linewidths and density structure in this and other
non-self-gravitating molecular clouds are due to the passage of a shock.  High-resolution CCS
maps of core D in TMC 1 in Taurus by Langer et al. (1995) show evidence for very small
condensations with sizes of order 0.01 pc (and narrow lines), and point out that these
condensations are unlikely to be the result of gravitational instability.  These observations
are consistent with the picture outlined in the present work, which provides a specific
mechanism for the production of internal velocity and density structure on a large range of
scales without the necessity of any sort of instability.

The timescale for conversion of the induced vortical motions into compressible modes should
be on the order of the cloud size divided by the induced vortical velocity for the model
shock-cloud system considered here.  However the present model is highly idealized, and in a
more realistic situation the shock will be corrugated and the cloud will contain pre-existing
density fluctuations, so the appropriate size scale and timescale should be smaller.

Schiano, Christiansen, \& Knerr (1995) performed two-dimensional simulations that are relevant to the
present work.  Although their study was concerned with high-speed winds interacting with clouds that
are initially ram-pressure confined, the results demonstrate the ability of shock-cloud interactions
to induce complex hierarchical internal density structures without disruption of the clouds. 
Particularly interesting is the fact that the derived perimeter-area fractal dimension of the
resulting internal density structure is in the range 1.3 to 1.4, very similar to that estimated
observationally for local Galactic density inhomogeneities.  Although Schiano et al. did not
explicitly discuss the induced internal vorticity and it is not clear how stratified the initial cloud
density distribution was, their results support our contention that the velocity field induced by the
shock-cloud interaction will quickly generate a strong compressional component which can account for
the observed density inhomogeneities over a large range of scales.

In the presence of a magnetic field, the induced vortical kinetic energy can also be channeled
into MHD waves.  However we forego any analysis of this transfer, except to note that the
Lorentz force can amplify vorticity as well as drain it (see Vazquez-Semadeni et al. 1996). 
We consider it reasonable to hypothesize that the vortical and compressible motions induced
by a shock can excite MHD waves to a degree that could explain the rough equipartition
observed between magnetic and kinetic energies in some clouds.  Thus the supersonic
linewidths might be due to MHD waves, but the source of excitation would be the shock
interactions examined here.  Given the need for such a source if supersonic linewidths are to
be attributed to MHD waves (Pudritz 1994, Stone 1994, MacLow et al. 1998, Ostriker et al. 1998),
transfer between these modes deserves a detailed examination.

It is important to note that the induced vortical and compressible velocity fields are
expected to be transformed into very different forms.  The vortical modes, being
supersonic, will easily be converted into compressible modes, and the purely fluid modes
couple to the magnetic field and, for the compressible modes, the gravitational field. 
Thus the observed vorticity in clouds will be much smaller than the initial value estimated
here.  The shocks create the initial energy that powers the turbulent or magnetic modes
which subsequently control the cloud evolution.  Thus, there is no inconsistency between
the large initial vortical motions calculated here  and the small values of shear or
rotation empirically observed in clouds (e.g. Goldsmith \& Sernyak 1984; Arquila \&
Goldsmith 1986; Goodman et al. 1993; Kane \& Clemens 1996).  We have only shown that the
shocks can provide the power to account for the energies in other, non-vortical, modes.

Finally, it is important to recognize that the shock-cloud interaction may result in gravitational
instability and fragmentation of the cloud, rather than simply keeping it ``stirred up."  First, some of the
density fluctuations induced by the vorticity could be gravitationally unstable.  Second, the shock itself
may become corrugated and generate condensations behind it, for example through the Vishniac instability
(Vishniac 1983, MacLow \& Norman 1993) if the shock can be regarded as pressure bounded on one side and if
the ratio of specific heats in the pre-shock gas is close enough to unity; however in this case the results
of MacLow \& Norman (1993) indicate that the postshock flows would be subsonic, and so unlikely to contribute
to the observed linewidths.  More likely, density and velocity fluctuations can be generated by the
interaction of the shock with the pre-existing turbulent velocity field, as in the calculations of
Kimura \& Tosa (1993) and Elmegreen et al. (1995).  Third, the larger scale compressive flattening of
the cloud, which occurs simultaneously with the vorticity generation, could promote gravitational
instability of the cloud, although this process is inhibited by the fact that the process studied in
this paper increases the velocity dispersion within the cloud as it undergoes flattening, making
instability more difficult.  Further consideration of these processes is clearly warranted, but is
beyond the scope of the present paper, which only aims at establishing the magnitude of the induced
increase of the cloud internal velocity dispersion.  Still, one must bear in mind that inhibition of
star formation by internal shock stirring is not the only possible outcome of the shock-cloud
interaction.

\section{SHOCK PUMP EFFICIENCY}
For shocks to be a significant source of internal cloud motions, the mean time between
exposures of a given cloud to a shock must be less than or comparable to the time for the
cloud to adjust to the previous shock by dissipation of the generated internal motions.  If
this condition is satisfied in general, then it seems unlikely that clouds can be in any
relaxed or equilibrium state.  A similar argument was used by Stone (1970) to argue that
cloud-cloud collisions are frequent enough to prevent clouds from attaining an equilibrium
configuration.  Here we neglect cloud-cloud collisions (which would only strengthen the
argument) and concentrate on exposures to supernovae (SN) and superbubbles (SB).  

For individual stars, the energy deposition by SN is expected to exceed that by other
individual stellar sources (winds from O stars, Wolf-Rayet stars, supergiants) by a
significant factor (e.g. Castor 1993), so we only include SN and SB.  If S is the rate per
unit volume of events that generate shocks and $R_{sh}(v_{sh})$ is the radius-velocity
relation for a particular class of events, then the frequency of exposures of a point in the
galactic disk to shocks of speed $v_{sh}$ or greater is, for a constant rate of events (Bykov
and Toptygin 1987; a more general expression for variable event rate is given by Ferriere
1992), the rate S multiplied by the volume of shells with speed $v_{sh}$.  Differentiating
this expression with respect to $v_{sh}$ gives the probability of the exposures to shocks of
speed in the range $v_{sh}$ to $v_{sh}+dv_{sh}$.  If the expansion law is
$R_{sh}=R_ov_{sh}^{-\alpha}$, the result is
\begin{equation}
P(v_{sh})=4\pi\alpha SR_o^3v_{sh}^{-(3\alpha+1)}
\end{equation}
\ni Using expressions given by Ferriere (1992; see also Cioffi and Shull 1991 for SN and
MacLow and McCray 1988 for SB), $\alpha\approx0.4$ for radiative SNRs, $\alpha=3/2$ for SBs
before they reach about 0.8 H (H = disk scale height) and $\alpha=1$ for SBs at larger
sizes.  This gives $-(3\alpha+1)$ in the range $-2.3$ (SN) to $-5.5$ (young SB), showing
that the vast majority of cloud exposures will be to shocks from shells traveling at their
lowest speeds, usually taken to be the general velocity dispersion of the ISM, $\sim5-10$
km/s.  (The same conclusion was reached independently by Heathcote and Brand
1983.)  Considering that the effective internal pressure of the clouds may have significant
contributions due to small scale turbulence and magnetic fields, this suggests that the most
important shock-cloud interactions will not be at very large Mach numbers (although the
actual appropriate values depend on one's view concerning the effective internal pressure),
and that most shock-cloud interactions will not disrupt the cloud.

The total exposure frequency can then be approximated as
\begin{equation}
\nu=\frac{4}{3}\ \pi R^3_{max}\ S
\end{equation}
\ni where $R_{max}$ is the maximum radius of the shell, at velocity $v_0$.  Using the
fiducial parameters adopted in the references given above, $R_{max}$ can be taken as about
70 n$_0^{-0.2}$ pc (n$_0$ = average density of the medium into which the SN or SB expands)
for SNRs for either bremstrahlung or metal cooling, and $R_{max}\approx400\ n_0^{-1/2}\ pc$
for SB (the n$_0$ dependence is $n_0^{-1/3}$ for older SBs; Ferriere 1992).  These values
assume the standard (and uncertain) values for SN energy (10$^{51}$ erg) and SB power source
(10$^{38}$ erg s$^{-1}$), but the dependence on these and other parameters is relatively
weak, and our general conclusions are not affected by the uncertainties.  We adopt the mean
event rates used by Ferriere (1994):  $S_{SNII}=1.3\times10^{-13}pc^{-3}yr^{-1}$ (the SNI rate
is much smaller), $S_{SB}=2.4\times10^{-15}pc^{-3}yr^{-1}$.  Then the mean time between
exposures, $\tau=1/\nu$, is
\begin{equation}
\tau_{SNII}=6\times10^6n_0^{0.6}\ yr
\end{equation}
\begin{equation}
\tau_{SB}=2\times10^6n_0^{1.2}\ yr
\end{equation}
\ni where the exponent 1.2 in $\tau_{SB}$ is a compromise between the values for older
(0.9) and younger ($\sim$1.5) superbubbles.  Notice that $n_0$ refers to the average
intercloud medium density in which the shocks propagate before they encounter the cloud,
not the cloud density.  The SBs dominate the exposure rate (at an average disk gas
density of $n_0=1\ cm^{-3}$) because their larger terminal volume outweighs their smaller rate
per unit volume.  Actually, since most SN II are expected to occur in clusters, and these
clustered SN contribute (along with OB star winds) to the superbubble rate, the timescale for
isolated SN may be largely irrelevant.

To estimate whether clouds have time to adjust to these exposure frequencies, there are
several alternatives.  1.  Following Stone (1970a,b), if adjustment is determined by the time
for rarefaction waves (whose speed is generally $\sim$3c, where c is the cloud internal
sound speed) to cross the cloud of size $\ell$ a few times, then the appropriate condition
for relaxation is $\tau\gta\ell/c$.  Using  the above expressions, this implies that only
clouds with sizes less than about 0.6 c$_{0.1}$ n$_0^{0.6}$ pc (SN II) or 0.2
c$_{0.1}$n$_0^{1.2}$ pc (SB), where c$_{0.1}$ is the internal sound speed in units of 0.1
km s$^{-1}$, will equilibrate quickly enough.  

2.  The mechanism discussed in this paper
generates vortical (and subsequent compressional) motions, and if these are identified with
the empirical linewidth-size scaling $\Delta v(\ell)\sim0.3\ell_{pc}^\alpha$, with
$\alpha\approx0.4-0.5$, the corresponding maximum cloud sizes for decay of induced motions
[with timescale assumed to be $\ell/\Delta v(\ell)$] are, for $\alpha=0.5,
\ \ell<4n_o^{1.2}$ pc (SN II) and $\ell<0.5n_o^{2.4}$ pc(SB).  We thus
conclude that shocks should provide an adequate ``pump" for internal cloud motions for
clouds larger than about 1 pc, if clouds could be thought of as independent entities
immersed in an intercloud or interclump medium of density not much more than 1 cm$^{-3}$.

3.  Finally, if the appropriate time for adjustment is the gravitational time
$\sim3\times10^7n_{cl}^{-0.5}yr\ (n_{cl}$ = internal cloud density), then clouds with
internal densities greater than about 25 $n_0^{-1.2}$ (SN II) or 220 $n_0^{-2.4}$ (SB)
could collapse before exposure to another shock.  However, since most small clouds are not
isolated, but are embedded within larger cloud structures, this estimate needs to be
repeated for SN and winds within GMCs, using rates as given by, for example, Bally et al.
(1991) and Miesch \& Bally (1994).  

As a simple example, consider the frequency of SN
shocks for a cluster of stars in a molecular cloud of average density $n_0$.  Assuming
that within the cloud complex $R^3_{max}$ scales as $n_0^{-0.6}$ and that the star formation
rate per unit volume scales linearly with the density, the results are as follows.  For
adjustment on the timescale of rarefactions (case 1 above), the maximum size for which there
is time for adjustment is $\ell_{pc}=0.6\ c_{0.1}n_0^{-0.4}$.  For the adjustment timescale
given by the velocity-size scaling (case 2 above), the maximum size is
$\ell_{pc}=4n_0^{-0.8}$.  So for densities characteristic of most observed molecular clouds
($n_0\sim10^2-10^4$), only very tiny clouds, with scales smaller than most current resolution
limits, will be able to escape encounters with shocks before they can equilibrate.  In the
case of gravitational collapse, the condition $\tau_{SN}>\tau_{ff}$ shows that clouds with
density enhancements greater than about 5 can collapse before they encounter another shock. 
However, since the preceding shock encounter is likely to generate internal motions (e.g. due
to MHD waves) that retard the collapse, the free-fall time is a strict lower limit, and
clouds with much larger density enhancement would be subjected to subsequent shocks before the
internal dissipation occurs.  It therefore appears that most clouds of observable scales will
be effectively pumped by shocks on a timescale less than their adjustment or collapse
timescale.  This argument for individual stellar sources within molecular clouds is much
stronger if one considers protostellar winds since, although the maximum expansion radius
is smaller, the rate per unit volume is much larger than for SN.

An independent comparison of the shock frequency (based on an evaporative expansion model
for supernovae), with the time for a cloud to re-establish presssure equilibrium, based on a
much more detailed model of the equilibration, has been given by Heathcote and Brand (1983). 
Their results agree well with the estimates given above, namely that clouds larger than
about 1 pc (depending on chosen parameters) will rarely have time to relax before being
overrun by another shock for Mach numbers less than about 10, and that the probability of
encounters with larger Mach number shocks is small.  Although they did not consider the
generation and decay of induced vortical motions, their results support our conclusion that
the shock pump is capable of sustaining the internal motions in clouds against the
dissipative decay of these flows.

The above estimates are very crude, and are only meant to be illustrative, since, for
example, explosion and wind events are highly clustered, ``clouds" are extremely irregular
and ill-defined and are probably hierarchically nested, etc.  However, the calculation
does suggest that the shock pump described here is capable of sustaining the internal
cloud motions that are observed.  The calculation furthermore suggests that it is unlikely
that clouds will be found in a state of equilibrium; some will be dispersing, some will be
contracting, and others may be in a temporary approximate state of virial equilibrium. 
The situation is similar to that found in the simulations of Vazquez-Semadini et al.
(1994), who suggest that it is simply the longer lifetimes of clouds that happen to be in
approximate virial equilibrium that is responsible for the fact that observations tend to
select them.

\section{SUMMARY AND IMPLICATIONS}
We have shown that shocks encountering model clouds with internal density gradients can
generate internal motions that are initially vortical and whose characteristic speeds are a
significant fraction of the shock speed and are comparable to both the compressive motions
directly excited by the shock and the bulk speed imparted to the cloud as a whole.  The
estimated induced velocity amplitudes and the scaling of the induced speeds with cloud
density contrast,
$v\sim\rho^{-1/2}$, are roughly consistent with observations, suggesting that the observed
linewidths are ultimately due to shock-induced velocity fields.  

In addition, we have shown
that the frequency of exposure of clouds to shocks is probably large enough that most clouds
are not in a state of equilibrium, but are instead constantly adjusting to the induced
velocity field due to the previously encountered shock, supporting the contention originally
made by Stone (1970a,b) in the context of cloud collisions. Only the smallest localized
structures can dissipate significant energy before encountering another shock. 

This initially vortical mode must
couple to the compressible modes and generate condensations within the cloud; the existence
of such condensations does not necessarily require any form of instability.  We pointed out a number of
observational studies of cloud morphology that reveal the presence of internal condensations which cannot
be attributed to either gravitational instability or internal protostellar winds.  We suggest that this
internal structure may be due to the shock-cloud interaction examined here.  Since the induced condensations
will be denser than and smaller than the cloud in which they are formed, the process may also produce an
inverse density-size scaling, although we do not know how to test this without detailed simulations.  More
importantly, the induced internal motions are supersonic, and so should generate further shock waves on
smaller scales, giving rise to a nested system of velocity and density fluctuations.  We suggest that this is
at least part of the physics behind the fractal-like structure observed in the cool ISM.  The vortical modes
will also excite MHD modes, and so, depending on the efficiency of the coupling, the results may still be
consistent with the observation of approximate equipartition between kinetic (linewidth) and magnetic energy
sometimes found in both self-gravitating and unbound clouds.  In this picture the ultimate source of the
linewidths is shock-induced velocity fields which both excite MHD waves and induce internal condensations.

An interesting problem concerns the scatter in internal velocity dispersion expected at a
given size.  In the present model this dispersion would be primarily due to the stochastic
variation in allowed decay times between successive shock encounters, in the Mach numbers of
incident shocks, and in the internal cloud properties before shock encounters.  We postpone
a discussion of the probability distribution of internal velocity dispersions to a
subsequent publication.  

The shock-induced motions studied here can provide a replenishment of supersonic turbulence
(in whatever form) for density fluctuations, or ``clouds," with sizes above a certain critical
scale.  This scale will not necessarily result in a ``signature" in the scaling relations between, for
example, cloud sizes and velocity dispersions, since in the present picture such scaling is due to
competition between turbulent dissipation and shock stirring, even at scales below the critical scale. 
Instead the critical scale is one below which velocity fluctuations have time to ``relax" and, perhaps, form
stars; it is only the internal structure that should show an imprint below this scale.  Although the
quantiative value of this scale must be fuzzy because of the existence of a distribution of cloud
properties, we tentatively identify this scale with the scale of ``velocity dispersion coherence" found by
Barranco et al. (1998).   The existence of such a (theoretical) scale may have important consequences relevant
to observations of the velocity fields in low-mass ``cores" and to the stellar initial mass function.

There is evidence that nearby ``cores" with sizes of order 0.1 pc have nonthermal internal
motions that are only slightly supersonic.  Furthermore it appears that the nonthermal
velocities are roughly independent of radius within a given core, unlike the decrease of
nonthermal linewidth with radius in larger structures, as demonstrated by Barranco and
Goodman (1997) and Goodman et al. (1997).  Based on these observations, and the result that
the velocity gradients within these cores appear decoupled from the gradients in the
environment, these authors suggest that the small cores are in a sense ``coherent" entities
in which gravity can overcome turbulence and lead to star formation.  Mapping at resolution
corresponding to
$\sim$0.01 pc will be necessary to determine whether or not they possess significant internal
density structures (see Langer et al. 1995 for convincing evidence that such density
substructure is present in the TMCI core D).  While they speculate that the phenomenon is due
to the increased effectiveness of ambipolar diffusion in small cores (see Myers and Goodman
1988b), the present work suggests an alternative interpretation, since it is just such
small-sized density fluctuations that should have sufficient time to dissipate much of their
turbulent energy before encountering another shock that would replenish the supersonic
motions.  According to this picture, the decrease of velocity dispersion with size among
clouds covering the entire range of observed size scales would be due to two effects:  1. The
dependence of induced internal velocities on density,
$\Delta v\sim\rho^{-1/2}$, since smaller regions are generally denser; 2. the larger waiting
times, relative to internal dissipation and/or adjustment timescales, for smaller clouds.  A
discussion of the resulting distribution of internal velocity fluctuations and their
dependence of size scale due to these two effects will be presented elsewhere (Scalo 2000, in
preparation).

If the clouds that are small enough to escape shock pumping for long enough to dissipate
significant turbulent energy are consequently able to form a star or stars, then this process
suggests a new theoretical interpretation of (one contribution to) the initial mass function:  
The probability that a region of a given mass will be able to escape shock pumping long 
enough to form a star should increase with decreasing size, and hence mass.  A realistic derivation of
the actual dependence of this probability on mass would be complex, for not only does it
involve the probability distributions of the shock velocities (which may be due to a number of
sources, including protostellar winds, supernovae, and wind-driven shells) and the density
field, but the density field itself is partially determined by the shock-induced motions on
different scales.  It seems unlikely that any analytic approach will be capable of estimating
the density fluctuation spectrum generated by the shock-induced vorticity and dilatation,
especially considering that MHD wave excitation will occur.  Although a simplified treatment
is possible (Scalo 1999, Scalo and Williams 2000), we postpone any calculation or comparison
with the observed IMF to a separate paper.  However it should be at least noted that the 
probability for escaping pumping long enough to collapse depends on the rate of arrival of
shocks for a given region of space, and this shock frequency is proportional to the average
star formation rate and IMF in the larger-scale neighborhood.  Thus the IMF model has an 
interesting feedback to the SFR itself, as can be seen in the simple example given in 
Scalo (1999).  For example, in galaxies without much star formation (either protogalaxies or
galaxies experiencing a lull), the shocks
may be primarily due to the supersonic turbulence itself, and as the turbulence decays, the shock
frequency  will decrease monotonically, greatly increasing the probability of shock pump escape.  The 
consequent increase in the SFR will boost the shock frequency, decreasing the escape 
probability, in turn modulating the SFR.

This ``escapist" or ``lucky cloud" scenario for the IMF is
in sharp contrast with suggestions that the IMF is controlled by the characteristic mass or mass
spectrum of a gravitational instability, or that star formation, through its associated shock wave
phenomena, triggers star formation rather than inhibits it, as in the present picture, or that a cloud
mass spectrum is set up by cloud collisions or instabilities, and that the mass of a star is then
determined by an interplay between accretion and outflow; Pudritz (1994) has discussed the
difficulties with the latter idea.  We suspect that all these processes, and more,
play a role.  However we do note that the C$^{18}$O study of the nearby L1333 region by Obayashi et al.
(1998) led them to suggest that star formation may occur preferentially in regions with smallest internal
kinetic energy (relative to self-gravitational energy), consistent with the ``lucky cloud" proposal.

The idea that the IMF, and star formation in general, is determined by a process in which
shocks related to young stars continually keep the neighborhood environment ``stirred up,"
except for the chance escape of local regions, preferentially small ones, for long enough for
dissipation to lead to another star formation event, is a specific example of a more general
inhibitory picture for star formation.  Young stars might inhibit star formation in
their neighborhood not just by the shock-induced internal motion discussed here but by thermal
inhibition by H II region heating, photoevaporation of clouds, disruptive cloud-cloud
collisions caused by the enhanced cloud-to-cloud velocity dispersion from bulk cloud
acceleration, and other processes.  Some of these processes will have their own distinctive
characteristic length scale below which escape from inhibition is more probable, similar to
the situation with the shocks discussed earlier, so that the IMF could have interesting
structure, although its theoretical calculation will be daunting.  We emphasize that the proposed inhibitory
contribution to the IMF is only one of several processes that may contribute to the form of the IMF (Scalo
1999).  In particular, as discussed in sec. 3.7 above, shock-cloud interactions may lead to {\it induced}
star formation for some range of shock velocities and cloud properties, and any comprehensive IMF theory will
have to take this process into account.  Conceptually, it is perhaps more useful to regard the ``lucky cloud"
conjecture as a modification of the ``spontaneous" star formation mode, a modification which should be
important in any interstellar medium which is supersonically turbulent, and hence dominated by shocks,
whether the shocks are generated by stellar energy input or by the turbulence itself.

Another interesting aspect of the stochastic shock pumping process described here concerns the
nature of the spatial organization of star formation that is possible in a given region,
whether on the scale of a cloud complex or of an entire galaxy.  Chappell and Scalo (1999)
examined a generic inhibitory model of this type, in which stellar ``stirring" or ``heating"
keeps neighboring regions from forming stars unless they can escape heating for a long enough
period that their velocity dispersion falls below a critical value.  They showed that such
models can exhibit several distinct ``phases" of spatial self-organization, including scattered
patches of star formation, oscillatory islands of star formation in a sea of steady star
formation, traveling waves of star formation, and even globally synchronized oscillations. 
Although these simulations are simplistic, they do illustrate that the theoretical
interpretation of the supersonic linewidths observed in cool localized regions of the
interstellar medium may have important consequences for understanding problems as diverse as
the initial mass function of stars and the spatial organization of star formation in
galaxies.

We are grateful to Robert Fleck for suggesting to one of us (JMS) the possible relevance of
the work of Chernin et al., and to Richard Klein for comments on an earlier version of
parts of this paper.  We also thank the referee, Bruce Elmegreen, for comments and critiques that improved the
presentation,  for pointing out the different expectations of morphology for sharp-edged model
clouds and the continuous density distributions considered here, and for reminding us of the potential
importance of induced star formation in contributing to the stellar initial mass function.  This work was
supported by NASA grant NAG 5-3107.

\clearpage

\begin{appendix}
\centerline{APPENDIX}
\vskip-14pt
\section{Effect of the Reflected Shock}

When a shock encounters a density {\it discontinuity}, a transmitted shock propagates into the
high density region, a reflected shock propagates back into the post-shocked gas, and a
contact discontinuity develops between them (see e.g. Landau and Lifshitz 1987).  The same
effect should occur when a shock encounters a continuous density gradient that increases in
the direction of shock propagation, which is the case considered in the present paper.  It is
therefore important to examine the ramifications of this effect, in particular the degree to
which it violates the assumption that the postshock pressure is constant along the shock line.

The pressure between the transmitted shock and the reflected shock must be greater than the
pressure behind the original shock in order to drive the reflected shock.  Further, the
pressure in the region between the two shocks depends on the density contrast (or gradient) in
a nonlinear fashion (Miesch and Zweibel 1994).  Because in general the density in front of the
shock may vary along the shock (i.e. perpendicular to the shock propagation), the
pressure and the Mach number of the shock may vary along the shock as well.  This pressure
gradient would interact with the density gradient to produce baroclinic vorticity.  In the
derivation of shock-generated vorticity given in the text, it was assumed that the pressure
behind the shock is constant along the shock.  This condition, for a discontinuity,
corresponds to a weak reflection or a weak density contrast.  For a continuous preshock
density distribution, we hypothesize that the condition is that the {\it gradient} be
shallow.  The results from simulations to be presented elsewhere support this hypothesis. 
Here we attempt an analytic treatment of the problem.

Let a shock with initial Mach number $M_s$ pass through some inhomogeneous medium with a
spatially varying density $\rho_0$.  Let
\begin{displaymath}
\delta_\tau\equiv\frac{\partial\rho_0}{\partial\tau}
\end{displaymath}
\ni be the derivative of the density parallel to the shock.  For the rest of this
Appendix the assumption is made that the results of Miesch and Zweibel (1994) for a shock
propagating into a density discontinuity can be used as a function of position for continuous
density gradients where the density contrast
\begin{displaymath}
\chi\sim\frac{\delta_n\ell}{\rho_0}
\end{displaymath}
\ni and $\ell$ is some length scale.  Since any variations in the Mach number and pressure
along the shock due to continuous density gradients are expected to be smaller than variations
due to a discontinuity, it is expected that the results in this section are upper limits.

According to Miesch and Zweibel (1994), for a strong ($M_s\gg1$) shock propagating into an
isobaric density {\it discontinuity} of density contrast $\chi$ and ratio of specific heats
$\gamma=5/3$, the trasmitted shock has a Mach number
\begin{equation}
M_t=2\left(\frac{\chi-\sqrt{1+15\chi/4}}{\chi-4}\right)M_s.
\end{equation}
\ni Differentiating this quantity along the shock yields
\begin{equation}
M_{t,\tau}=\frac{2\chi_\tau}{(\chi-4)^2}\left[(\chi-4)\left(1-\frac{15/8}{\sqrt{1+15\chi/4}}
\right)-\chi+\sqrt{1+15\chi/4}\right]M_s,
\end{equation}
\ni where
\begin{displaymath}
\chi_\tau\equiv\frac{\partial\chi}{\partial_\tau}=\frac{\delta_{n,\tau}\ell}{\rho_0}-
\frac{\delta_n\delta_\tau\ell}{\rho^2_0}.
\end{displaymath}
\ni In the limit of large $\chi$, equation (A2) reduces to
\begin{equation}
M_{t,\tau}\sim\frac{\sqrt{15/4}\chi_\tau M_s}{\chi^{3/2}}.
\end{equation}

For a strong shock with a ratio of specific heats $\gamma=1$, the Mach number of the
transmitted shock is given by
\begin{equation}
M_t=\frac{\sqrt{\chi}M^2_s}{M_s+\chi}.
\end{equation}
\ni The derivative along the shock is
\begin{eqnarray}
\nonumber M_{t,\tau}&=&
\frac{\chi_\tau}{(M_s+\chi)^2}\left[\frac{(M_s+\chi)M^2_s}{\sqrt{\chi}}-M^2_s\right] \\
&=& \frac{M^2_s\chi_\tau}{\sqrt{\chi}(M_s+\sqrt{\chi})^2}.
\end{eqnarray}

Eqs. A3 and A5 show that the logarithmic gradient of the Mach number along the shock line,
$\partial\ell nM/\partial\tau$ is smaller than the logarithmic gradient of the density along
the shock line, $\partial\ell n\chi/\partial\tau$, by a factor which is about $\chi^{-1/2}$
for
$\gamma=5/3$, $\chi^{-1/2}$ for $\gamma=1$ when $\chi^{1/2}\gg M_s$, and $\chi^{1/2}/M_s$ for
$\gamma=1$ when $M_s\gg\chi^{1/2}$.  In other words, even though the length scale for density
variations along the shock is of order the cloud size (by the definition of ``cloud"), the
scale for variation of the Mach number is much larger, at least for large $\chi$ and Mach
number, giving approximate support for the assumption of constant Mach number along the shock
made in sec. 2.4. of the text.  The case in which $\chi$ and/or $M$ are not large is
considerably more complicated, especially since dynamic postshock vorticity generated by the
downstream gradients is not negligible in that case.
\end{appendix}

\end{document}